\documentclass[useAMS,usenatbib]{mn2e}
\usepackage[dvips]{graphicx}
\usepackage{amsmath}
\usepackage{amssymb}
\usepackage{amsfonts}
\usepackage{aas_macros}
\usepackage{ulem}

\title[Rapidly Rotating Lenses] {Rapidly Rotating Lenses: Repeating
  features in the lightcurves of short period binary microlenses}
\author[M. T. Penny, E. Kerins and S. Mao]
  {Matthew~T.~Penny,$^1$\thanks{mpenny@jb.man.ac.uk}  Eamonn
    Kerins,$^1$ Shude Mao$^{2,1}$ \\
$^1$Jodrell Bank Centre for Astrophysics, The Alan Turing Building,
  School of Physics and Astronomy,\\ The University of Manchester,
  Oxford Rd, Manchester, M13 9PL, UK\\
$^2$National Astronomical Observatories, Chinese Academy of Sciences, A20 Datun
Road, Chaoyang District, Beijing 100012, China}
\date{Accepted 2011 July 7. Received 2011 July 5; in original form 2011 May 7}

\def\LaTeX{L\kern-.36em\raise.3ex\hbox{a}\kern-.15em
    T\kern-.1667em\lower.7ex\hbox{E}\kern-.125emX}

\begin{document}

\newcommand{\sdet}{s_{\rm det}}
\newcommand{\contwofac}{2.85\text{~AU~}}
\newcommand{\threshold}{\Delta m_{\mathrm{min}}}
\newcommand{\uzeromax}{u_{0,\mathrm{max}}}
\newcommand{\TtEconst}{4.51\text{~AU~}}
\newcommand{\detectability}{\varepsilon}
\newcommand{\fblend}{f_{\mathrm{s}}}
\newcommand{\psif}{\psi_{\mathrm{f}}}
\newcommand{\tf}{t_{\mathrm{f}}}
\newcommand{\phizero}{\phi_0}
\newcommand{\mbase}{m_{\mathrm{b}}}
\newcommand{\paczynski}{Paczy{\'n}ski }
\newcommand{\besancon}{Besan{\c c}on }
\newcommand{\zs}{z_{\mathrm{s}}}
\newcommand{\zone}{z_{\mathrm{1}}}
\newcommand{\ztwo}{z_{\mathrm{2}}}
\newcommand{\zonebar}{\overline{z}_{\mathrm{1}}}
\newcommand{\ztwobar}{\overline{z}_{\mathrm{2}}}
\newcommand{\zbar}{\overline{z}}
\newcommand{\bigmone}{M_{\mathrm{1}}}
\newcommand{\bigmtwo}{M_{\mathrm{2}}}
\newcommand{\vt}{v_{\mathrm{t}}}
\newcommand{\blendfs}{f_{\mathrm{s}}}
\newcommand{\uzero}{u_{\mathrm{0}}}
\newcommand{\tzero}{t_{\mathrm{0}}}
\newcommand{\tein}{t_{\mathrm{E}}}
\newcommand{\thetae}{\theta_{\mathrm{E}}}
\newcommand{\re}{r_{\mathrm{E}}}
\newcommand{\dl}{D_{\mathrm{l}}}
\newcommand{\ds}{D_{\mathrm{s}}}
\newcommand{\kms}{km~s$^{-1}$}
\newcommand{\msun}{M_{\sun}}
\newcommand{\mjup}{M_{\mathrm{Jupiter}}}
\newcommand{\rsun}{R_{\sun}}
\newcommand{\pie}{\pi_{\mathrm{E}}}
\newcommand{\numpi}{\mathrm{pi}}

\label{firstpage}

\maketitle

\begin{abstract}
Microlensing is most sensitive to binary lenses with relatively large
orbital separations, and as such, typical binary microlensing events show
little or no orbital motion during the event. However, despite the
strength of binary microlensing features falling off rapidly as the
lens separation decreases, we show that it is possible to detect repeating
features in the lightcurve of binary microlenses that complete several
orbits during the microlensing event. We investigate the lightcurve
features of such Rapidly Rotating Lens (RRL) events. We derive
analytical limits on the range of parameters where these effects are
detectable, and confirm these numerically. Using a population
synthesis Galactic model we estimate the RRL event rate for a
ground-based and space-based microlensing survey to be $0.32
f_{\mathrm{b}}$ and $7.8 f_{\mathrm{b}}$ events per year respectively,
assuming year-round monitoring and where $f_{\mathrm{b}}$ is the binary fraction. We detail how RRL event parameters can
be quickly estimated from their lightcurves, and suggest a method to
model RRL events using timing measurements of lightcurve
features. Modelling RRL lightcurves will yield the lens orbital period and
possibly measurements of all orbital elements including the
inclination and eccentricity. Measurement of the period from the
lightcurve allows a mass-distance relation to be defined, which when combined
with a measurement of microlens parallax or finite-source effects, can
yield a mass measurement to a two-fold degeneracy. With sub-percent
accuracy photometry it is possible to detect planetary companions, but
the likelihood of this is very small.
\end{abstract}


\section{Introduction}
\label{Introduction}

By monitoring hundreds of millions of stars towards the Galactic bulge
and Magellanic clouds, Gravitational microlensing surveys such as
OGLE~\citep{Udalski:2003ews} and MOA~\citep{Hearnshaw:2006moa} detect
$\sim 1000$ microlensing events per year. The lightcurves of most
microlensing events follow the typical \citet{Paczynski:1986gml} form
of a point-mass lens with a point source. However, many
event lightcurves have a more complex form due to the effects of a
binary or planetary companion to the
lens~\citep{Mao:1991bml,Gould:1992pmm}, a binary companion to the source~\citep{Griest:1992bsm}, microlens parallax~\citep{Refsdal:1966lmd,Gould:1992mss}, finite-source size~\citep{Gould:1994mte,Witt:1994fs,Nemiroff:1994mfs}, or a combination of such. These effects provide
additional information about the lens, such as the mass ratio and
projected separation in binary or planetary lens events, or
mass-distance relationships with parallax or finite-source
effects. Measurement of both the finite-source size and
microlensing parallax allow the lens mass to be solved for
uniquely~\citep{Gould:1992mss}, and if these occur in a binary or
planetary lensing event, also allow measurement of the companion
mass.

The complexities of microlensing lightcurves, to a greater or lesser extent, can all be considered as
deviations from the single lens \paczynski
form. The deviations may be relatively minor
and can cover the entire lightcurve, as in most parallax
events~\citep[e.g.][]{Smith:2002ope}, or they can be large and cover
only a small fraction of the lightcurve, as in many binary lens events~\citep[e.g.][]{Kubas:2005blo,Beaulieu:2006fem}. In binary
lens events, these deviations from the single lens form are caused by
a difference in the magnification pattern of the lens. The most
prominent `features' of the binary lens magnification pattern are
caustics, where the magnification of a point source diverges (see Figure~\ref{exampleMagMap}). A source
passing over a caustic will show a sharp rise in its magnification as
it enters it and a sharp fall as it leaves. Other, more smooth
magnification pattern `features' can be associated with the caustics
in some way.

The caustic features of a binary lens magnification pattern are a
natural feature of the binary lens mapping
\begin{equation}
\zs = z - \frac{1}{1+q}\left( \frac{1}{\zbar - \zonebar} + \frac{q}{\zbar - \ztwobar}\right),
\label{lensMapping}
\end{equation}
which maps the angular position of the source to image positions under the
influence of the lens, and where we have used complex coordinates
\citep[e.g. $z = x + iy$,][]{Witt:1990blc}; bars represent complex
conjugation, $\zs$ is the position of the source, $z$ the position
of the image, $\zone$ and $\ztwo$ the positions of the primary and
secondary lens components respectively, and $q=\bigmtwo/\bigmone$ is
the mass ratio of the lens components. All angles have been normalized
to the Einstein ring radius
\begin{equation}
\thetae = \frac{\re}{\dl} = \frac{1}{\dl}\sqrt{\frac{4G}{c^2}x(1-x)\ds
  M},
\label{einsteinRadius}
\end{equation}
where $\re$ is the physical Einstein radius, $\dl$ and $\ds$ the
distance to the lens and source respectively, $M=\bigmone+\bigmtwo$ is
the total lens mass, $x=\dl/\ds$ is the fractional lens distance, and
$G$ and $c$ are the gravitational constant and speed of light
respectively. The magnification $\mu$ of an image is given by the
determinant of the Jacobian of the lens mapping
\begin{equation}
\mu = \frac{1}{\det J}.
\end{equation}
The magnification diverges when $\det J = 0$, and the solutions of
this equation form smooth, closed curves in the image plane called
critical curves, which when mapped back to the source plane form
closed, cuspy curves: the caustics (see Figures~\ref{exampleMagMap}
and \ref{parametrization}).

In a binary lens event, the caustics are largest when the projected
lens separation $s = |\ztwo - \zone| \sim 1$, i.e., the lens components
orbit with semimajor axis $a \sim \re \sim 2$--$3$~au. At these
separations there is only a single, so-called resonant,
caustic. Lenses with wider separations have two caustics, while those
with closer separations have three~\citep{Schneider:1986bgl}. The caustics
of close and wide systems have smaller caustics than resonant systems,
and so the source is less likely to encounter them. Far from the
caustics, the binary lens features tend to be weak, and therefore, the
lightcurves of binary lenses with very close or very wide orbits
(which have correspondingly very small caustics), in most cases, will
be indistinguishable from those of a single lens~\citep[e.g.][]{Gaudi:1997cbm}.

The binary lenses with the strongest lightcurve features, thus have
orbital periods $T \sim 1000$~d, much longer than the microlensing
event timescale 
\begin{equation}
\tein = \frac{\re}{\vt} \sim 20 \text{d},
\label{einsteinTimescale}
\end{equation}
for a typical Galactic microlensing event, where $\vt$ is the relative
lens-source transverse velocity (source velocity). The lenses
therefore complete only a small fraction of their orbit during the
course of the microlensing event, and only a fraction of the events
are expected to show detectable signs of orbital motion in their
lightcurves~\citep{Gaudi:1997cbm,Dominik:1998mrb, Konno:1999aml, Ioka:1999kre, Rattenbury:2002hmp, Penny:2011omm}. Those events where it has been detected typically
involve caustic crossings, because the sharp caustic crossing features
on the lightcurve allow precise timing, which can be used to constrain
even small lens motions~\citep{Albrow:2000rbl, An:2002eb5,
  Gaudi:2008jsa, Ryu:2010drb, Hwang:2010vlm, Skowron:2011kos,
  Batista:2011mdp}. In these events, the orbital motion has
constrained the properties of the lens orbit, and in one case allowed
the measurement of several orbital
parameters~\citep{Bennett:2010jsa}. 

In a previous paper, we have shown that with accurate photometry and dense
enough lightcurve coverage, it is possible to detect orbital motion in
lenses with closer orbits, and weaker, smooth lightcurve
features~\citep{Penny:2011omm}. Typically occurring in events with
longer microlensing timescales, $\tein \sim 100$~d, the shorter
orbital periods means that the lens completes a much larger fraction
of the orbit during the event. In this paper we investigate the extreme
case, where the lens completes at least one orbit during the
microlensing event. In events involving such rapidly rotating lenses,
magnification pattern features can sweep over the source more than
once, and if detected in the lightcurve, these repeating features
allow an accurate measurement of the lens period. Knowledge of the
period places constraints on the lens mass, and if combined with a
measurement of the finite-source effect or parallax, can be used to
measure the lens mass to a two-fold degeneracy~\citep{Dominik:1998mrb}.

We begin the paper by defining when a binary lens is a rapidly
rotating lens in Section~\ref{RRLs}. In Section~\ref{detectability} we
look at whether rapidly rotating lenses are detectable, and the rate
at which they are expected to occur. In
Section~\ref{physicalParameters}, we then look at what information can
be gained from observing a rapidly rotating lens, and suggest a method
for modelling such events. In two appendices we examine how rapidly
rotating lenses affect the microlensed images, and introduce additional
effects that can affect rapidly rotating lens lightcurves.

\section{What is a Rapidly Rotating Lens?}
\label{RRLs}

\begin{figure}
\includegraphics[width=84mm]{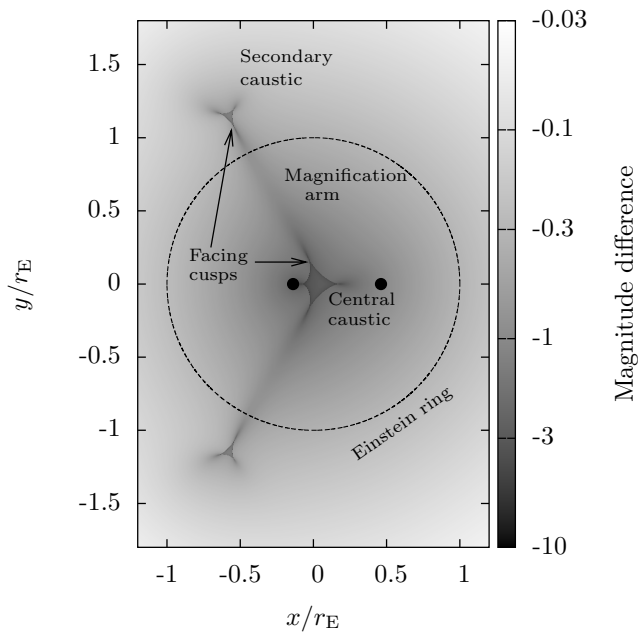}
\caption{The magnification pattern of a close topology microlens. The
  dots denote the lens positions, with the primary lens at
  negative $x$. The lens has a mass ratio $q=0.3$ and projected
  separation $s=0.6$. Notable features of the magnification
  pattern are labelled.}
\label{exampleMagMap}
\end{figure}

We define a rapidly rotating lens (RRL) to be a binary microlens, which
if monitored continuously with suitable photometric accuracy, would
guarantee that at least one feature of its magnification pattern would
be seen to repeat at least once in its lightcurve, due to the lens
orbital motion. This implies that the lens completes at least two
orbits during the time in which its binary lensing features are
detectable. We choose this definition over the more simple comparison
of microlensing and orbital timescales (e.g. $T < \tein$) because
without detecting binary features it is impossible to measure the
binary's rotation. As mentioned in the previous section, the strength
of binary features declines as the orbital separation and period
decrease, so simply decreasing the period does not necessarily
increase the prospects of detecting a repeated feature. A rapidly
rotating lens must therefore compromise between a fast rotation rate
and detectable binary lensing features.

Throughout the paper we shall focus on close topology lenses, which
have separations $s\lesssim 0.7$~\citep{Schneider:1986bgl,Erdl:1993ctc}, a choice we
shall justify in Section~\ref{detectability}. Figure~\ref{exampleMagMap}
shows the magnification pattern of a close topology lens, and labels a
number of features. The structure and features of the magnification
pattern depend only on the projected separation of the lens components
$s$, and the mass ratio $q$~\citep{Erdl:1993ctc}. The most important
features of the close magnification pattern are a central caustic,
located at the lens centre of mass, and two secondary caustics which
lie away from the lens centre. Stretched between the central and secondary
caustics are two `arms' of excess magnification (relative to the
magnification that would be caused by a single lens of mass equal to
the total binary mass). During a microlensing event, a source will
travel across the magnification pattern, and we will observe the
source to change in brightness, the form of this lightcurve being
determined by the trajectory that the source takes. As the source
moves, the magnification pattern will not stay fixed, as the binary
will also move in its orbit. Should the lens orbit lie face-on to the
line of sight, then the magnification pattern will rotate as the
source moves across it. Should the orbit be inclined or eccentric, the
structure of the magnification pattern will also change, as it depends
on the projected lens separation $s$~\citep{Schneider:1986bgl}.

\begin{figure}
\includegraphics[width=84mm]{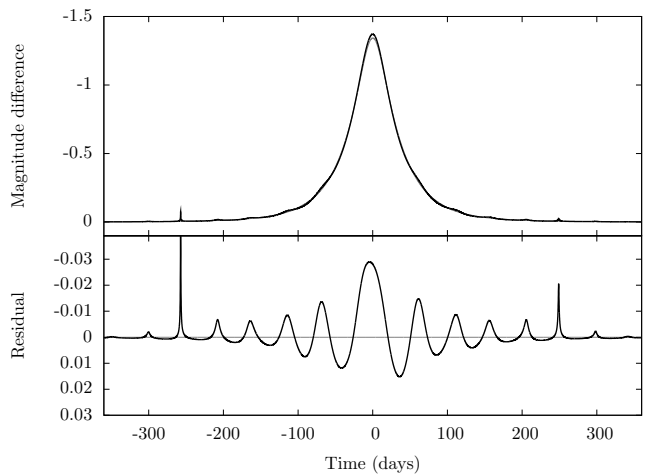}
\caption{The lightcurve of a rapidly rotating lens. The upper panel
  shows the RRL lightcurve in black, and the \paczynski lightcurve of a
  single lens with the same total mass in grey. The lower panel shows
  the residual with respect to the \paczynski lens
  lightcurve. Features due to the magnification arms appear as peaks in
  the residual, while between them there are relative
  demagnifications. Large, short duration spikes occur when the
  secondary caustic passes close to or over the source. The system has
  parameters $\tein=61$~d, $T=92$~d, $s=0.23$, $q=0.8$, $\uzero=0.3$,
  $\phizero=1.75$ (see Section~\ref{physicalParameters} for
  definitions of $\uzero$ and $\phizero$).}
\label{exampleLightcurve}
\end{figure}

Figure~\ref{exampleLightcurve} shows the lightcurve (thick black line) of a
RRL with a similar magnification pattern to that shown in
Figure~\ref{exampleMagMap}. It closely resembles the lightcurve of a
single lens, the \paczynski lightcurve (grey line), but with a
quasi-periodic variation over the entire lightcurve that only becomes
obvious in the residual that is left once the \paczynski curve is
subtracted from the lightcurve. These periodic features correspond to the magnification arms that extend between the secondary and central caustics, which sweep over
the source as the lens rotates. The microlensing timescale of the
lightcurve shown is $\tein=60$~d, but it is clear that repeating
binary features remain in the lightcurve at time from peak
magnification much greater than this, which corresponds to a source
position far outside the Einstein ring. This is because the secondary
caustics can lie far outside the Einstein ring, their distance from
the lens centre increasing as the binary separation
decreases. However, their size, and the strength of the magnification
arm connecting them with the central caustic, decrease with decreasing
binary separation. We note at this point, that despite the
  large separation of the secondary caustics, we need not consider
  relativistic effects of superluminal caustics~\citep{Zheng:2000slc}
  as the ratio of the caustic rotational speed to the speed of light
  in all the cases we will consider is $\sim 10^{-3}$.

An RRL can clearly exhibit interesting, repeating lightcurve features if the
binary period and separation conspire, but until now we have only
considered RRLs in relation to parameters normalized to the typical
microlensing length- and time-scales. In order to see if RRL events
will be detectable in real microlensing surveys we must consider how
their properties relate to the physical parameters of the lensing system.

\section{Are RRLs detectable?}
\label{detectability}

In the previous section we defined a criterion for a lens to be a
rapidly rotating lens, and described the features of an RRL event. In
this section we put the definition on a more quantitative basis, and
investigate whether RRLs will occur amongst the microlenses that are
probed by microlensing surveys. To determine if detection is
plausible, we investigate the range of physical parameters required to
produce a microlensing event with repeating features, first
analytically and then numerically. Finally we apply our numerical
method to simulated microlensing surveys to estimate the expected rate
of RRL detections.

\subsection{An analytical approach}
\label{analyticalDetectability}

To see repeating features in a microlensing event, the most
fundamental requirement of the system is that the lens completes more
than one orbit during the event. The magnification pattern of a
binary lens is complicated, but the essential features of a close
binary lens can be captured by assuming it to be composed of two
straight, radial arms that extend from the centre of mass to the
position of the secondary caustics. Under this assumption, and
assuming a random initial phase angle, repeating features are
guaranteed to be observed if the lens completes two orbits in the time
that the source spends within the radius swept out by the arms. We can write this as an inequality
\begin{equation}
2T < \frac{\upi}{2}u_{\pm}\tein,
\label{TtE_limit}
\end{equation}
where $u_{\pm}$ is the radial position of the secondary caustics in
units of Einstein radii (see Figure~\ref{parametrization}), and the factor of $\smash{\frac{\upi}{2}}$ is
the mean chord length across a unit circle, and accounts for the random
impact parameter of source trajectories relative to the lens centre of
mass. It should be noted that it is possible for a feature to repeat
if the binary completes between one and two orbits, but this
requires a coincidence in the timing of the first feature. 

Both the orbital period and the Einstein timescale depend on the lens
mass, and the period also depends on the lens semimajor axis, so it is
possible to write this constraint in terms of $M$ and $a$. For
projected lens separations $s\ll 1$, \citet{Bozza:2000cmc} has derived
an analytical approximation for the secondary caustic positions (see Equation~\ref{causticPosition}), which
if we keep only the first order terms is
\begin{equation}
u_{\pm}(s,q) \simeq s^{-1}.
\label{causticRadialPosition}
\end{equation}
Using equations~\ref{einsteinRadius} and \ref{einsteinTimescale} and
Kepler's third law, with a little algebra we can then write
Equation~\ref{TtE_limit} as a constraint on the semimajor axis of the binary
\begin{equation}
a < \TtEconst [x(1-x)]^{2/5} \ds^{2/5}
\vt^{-2/5} M^{3/5},
\label{upperLimit}
\end{equation}
where we have assumed a face-on orbit so that $s=a/\re$, and where $M$
is the total lens mass in solar masses, $\ds$ the source distance in
kpc, $x\equiv \dl/\ds$ is the ratio of lens and source distances and
$\vt$ the relative lens-source velocity in \kms.

While we have an upper limit on the lens semimajor axis, in order to
detect the RRL lightcurve features, they must be strong enough to be
detectable in the photometry of the microlensing event. This
requirement is somewhat ambiguous, but as the magnification pattern
depends only on $s$ and $q$, and the strength of features decreases
with decreasing $s$, we can assume that for a given photometric
precision and mass ratio, magnification pattern features will be
detectable only when the separation is larger than a certain value, i.e.
\begin{equation}
s > \sdet,
\label{constraint2}
\end{equation}
where $\sdet$ depends on $q$ and the photometric accuracy. For stellar
binary mass ratios, there will only be a small dependence on $q$, but
there will be a strong dependence on the photometric accuracy; a value
of $\sdet=0.3$ is reasonable, however (see
Section~\ref{numericalDetectability}). We can again write this
constraint as a limit on the semimajor axis
\begin{equation}
a > \contwofac \sdet [x(1-x)]^{1/2} \ds^{1/2} M^{1/2}.
\label{lowerLimit}
\end{equation}

\begin{figure}
\includegraphics[width=84mm]{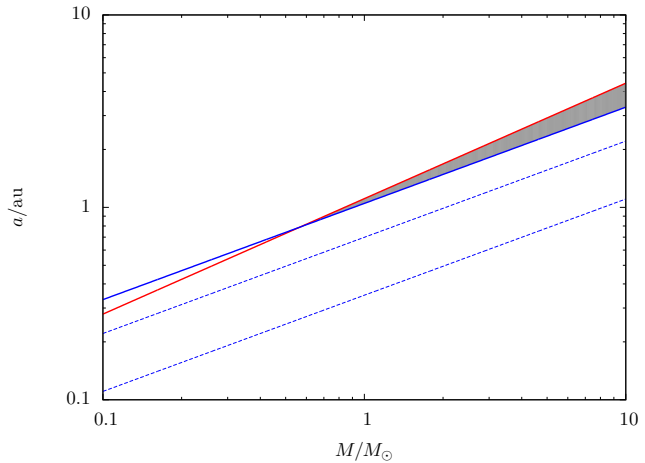}
\caption{Plot showing the region of the total mass-semimajor axis
  plane where repeating features are observable. The red line shows the
  upper limit on $a$, provided by the constraint in
  Equation~\ref{upperLimit}, while blue lines show the lower limit on
  $a$, provided by the constraint in Equation~\ref{lowerLimit}, with values of
  $\sdet=0.3, 0.2$ and $0.1$ from top to bottom. The other parameters
  are set at $x=0.75$, $\ds=8$~kpc and $\vt=50$~\kms. The region where
  repeating features are detectable for $\sdet=0.3$ is shaded grey.}
\label{analyticalConstraints}
\end{figure}

We now have two constraints on $a$, an upper and a lower limit, which
are dependent on other parameters of the lensing system, the most
interesting being the total lens
mass. Figure~\ref{analyticalConstraints} shows the two constraints on
semimajor axis as a function of mass, for a lens system with $x=0.75$,
$\ds=8$~kpc and $\vt=50$~\kms, with values of $\sdet=0.3, 0.2$ and
$0.1$. Other than the slow lens-source velocity ($\langle \vt \rangle
\approx 200$~\kms{} for a Bulge microlensing event), these values are
typical of a microlensing event towards the Galactic Bulge. The plot
shows that most of the parameter space is excluded, but thanks to the
differing power-law indices on the mass dependence, there is a small
range of parameters over which repeating features should be
detectable. For the parameters shown, the `detectable region' opens up
at $M \sim 1\msun$ and $a \sim 1$~au, and widens to $3.3 <
a/\text{au} < 4.4$ by $M=10\msun$. The dependence of the limits on
other parameters means that the region of detectability will get
smaller, and move to larger $a$ as the source distance grows, will get
larger and move to smaller $a$ and $M$ as the lens moves closer to the
source or the observer, and will get smaller as the relative
lens-source velocity increases. A small but significant fraction of
binary stars will have total masses and semimajor axes in the range of
detectability~\citep[e.g.][]{Duquennoy:1991bsp}, especially if
improved photometric accuracy can reduce $\sdet$.

\subsection{A numerical approach}
\label{numericalDetectability}

In deriving analytical limits on the range of lens parameters we have
had to make assumptions about the magnification pattern and strength
of features. If we instead proceed numerically, we need not make these
assumptions as we can determine precisely the regions of the
magnification pattern where features are detectable for any given
photometric accuracy. We define a detectability $\detectability$ that
is the probability that, for a given lens system and photometric
precision, an RRL with a face-on orbit will exhibit at least one
detectable repeating feature in its lightcurve. A feature is said to
be detectable at a radial position $u$, if the range of magnifications $\mu$
over a circle of radius $u$ satisfies
\begin{equation}
\Delta m \equiv 2.5 \log \left[ \frac{\mu_{\mathrm{max}}(u)}{\mu_{\mathrm{min}}(u)} \right] \ge \threshold,
 \label{detectabilityeq}
\end{equation}
where we have expressed the range of magnifications ($\mu_{\mathrm{min}} \rightarrow \mu_{\mathrm{max}}$) as a magnitude
difference $\Delta m$, and $\threshold$ is the photometric detection threshold,
which can be taken to mean the typical uncertainty in magnitude on a
data point in the baseline of the lightcurve. In calculating
$\detectability$ we average over the random parameters of the source
trajectory and phase angle.

\begin{figure}
\includegraphics[width=84mm]{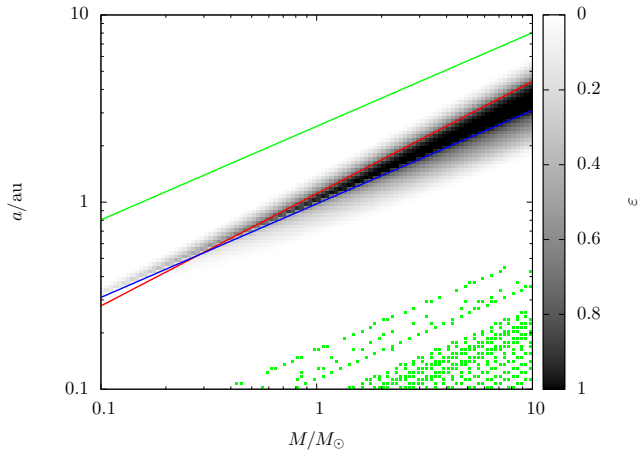}
\caption{Plot of the average detectability $\detectability$ against
  total lens mass $M$ and semimajor axis $a$, for a lens with mass
  ratio $q=0.3$. The lens and source distances and relative velocity
  are the same as in Figure~\ref{analyticalConstraints}. The red line is the
  analytical upper limit of Equation~\ref{upperLimit}, while the blue
  line is the analytical lower limit of Equation~\ref{lowerLimit} with
  a value of $\sdet=0.28$ for a photometric precision
  $\threshold=0.01$. The green line at the top of the figure marks the
  boundary between regions of close and resonant topology lenses -- we
  only calculate $\detectability$ for close topology lenses. The green
  points lower in the figure mark points where our calculation of
  $\detectability$ failed (see text for details).}
\label{numericalConstraints}
\end{figure}

We can now test the predictions we made in
Section~\ref{analyticalDetectability}, by mapping the
detectability $\detectability$ against total mass $M$ and semimajor
axis $a$, for the set of parameters we used for
Figure~\ref{analyticalConstraints}. Figure~\ref{numericalConstraints}
shows regions of finite detectability in shades of gray up to black
when $\detectability=1$. Green points in the plot show where the
calculation of the detectability, which requires several numerical
minimization and root finding steps, failed. The green line at the top
shows the boundary between the close and resonant caustic regimes,
where the three caustics of the close topology merge into a single
resonant caustic. We do not calculate the detectability in the
resonant regime. The red and blue lines show the analytical upper and
lower limits of Equations~\ref{upperLimit} and \ref{lowerLimit},
however with $\sdet=0.28$ as opposed to $0.3$. It can be seen in the
figure that the analytical upper limit of Equation~\ref{upperLimit}
agrees very well with the numerical region of detectability,
coinciding with the boundary where $\detectability$ begins to fall
from unity with increasing $a$ over the entire range of $M$ shown. Equation~\ref{upperLimit}, without the factor of $2$ that was introduced on the left-hand side of Equation~\ref{TtE_limit} to guarantee a repeated feature, also describes well the region where detection becomes possible but is not guaranteed
(i.e. $0<\detectability<1$). 

The analytical lower limit, once the parameter $\sdet$ has been
adjusted to $0.28$ for a guaranteed repeating feature, also agrees
well with the numerical region of detectability. It should be noted
however, that the slope of the lower edge of the numerical
region is slightly shallower than the analytical lower limit. This
becomes more pronounced when the lens gets closer to the source,
the total mass increases or the source velocity decreases. This is
because the assumption, that there are detectable features over the
entire magnification pattern within $u<u_{\pm}$, breaks
down, and the detectable features lie in two disjoint regions, a disc
surrounding the central caustic and an annulus containing the
secondary caustics. The size of these regions depends on $s$, and the
lower limit on $a$ (Equation~\ref{lowerLimit}) becomes a shallower
function of $M$. This effect is more important in determining the
slope of the lower limit on $a$ where $\detectability=0$.

\begin{figure*}
\includegraphics[width=84mm]{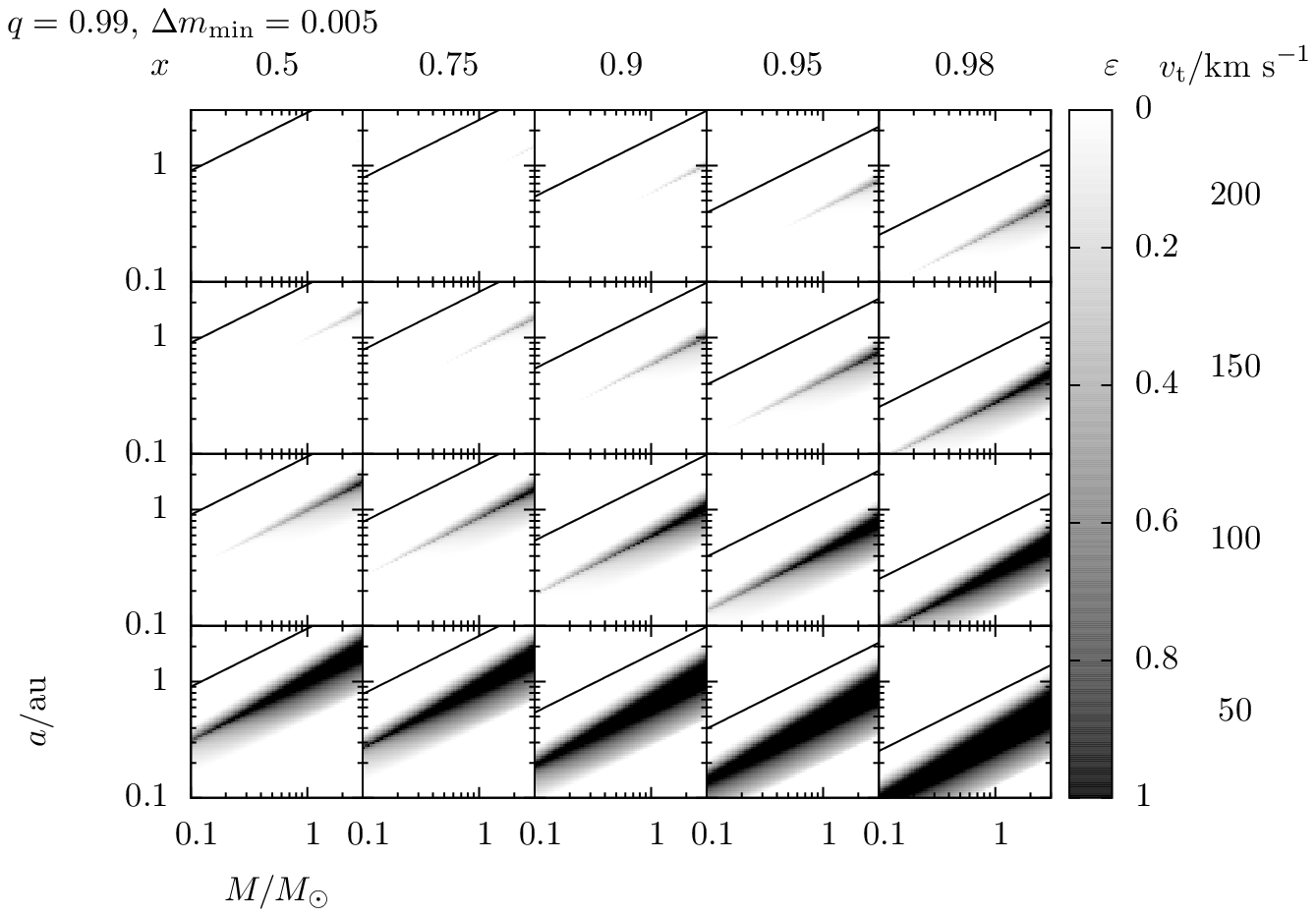}
\hspace{0.02\textwidth}
\includegraphics[width=84mm]{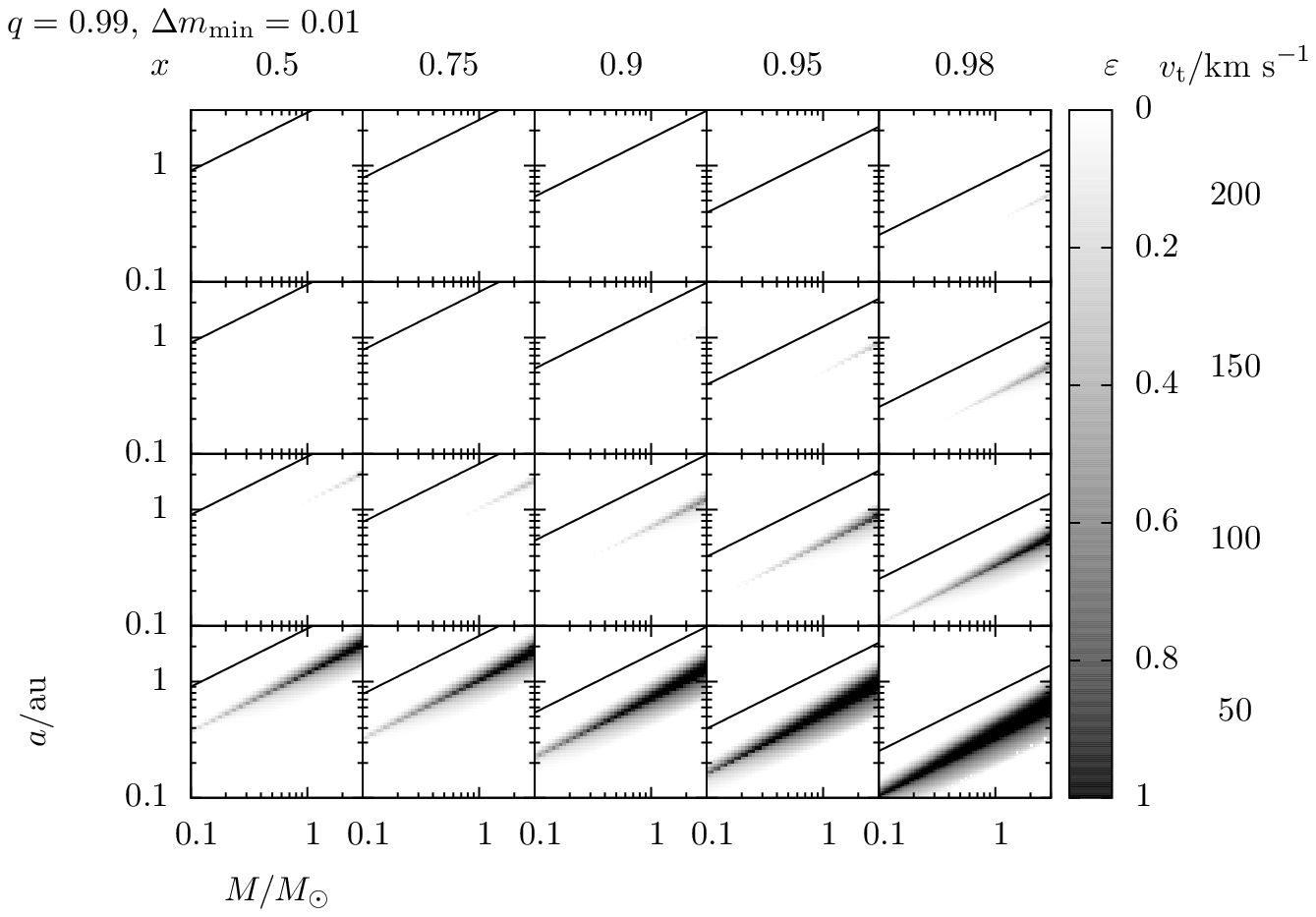}\\
\includegraphics[width=84mm]{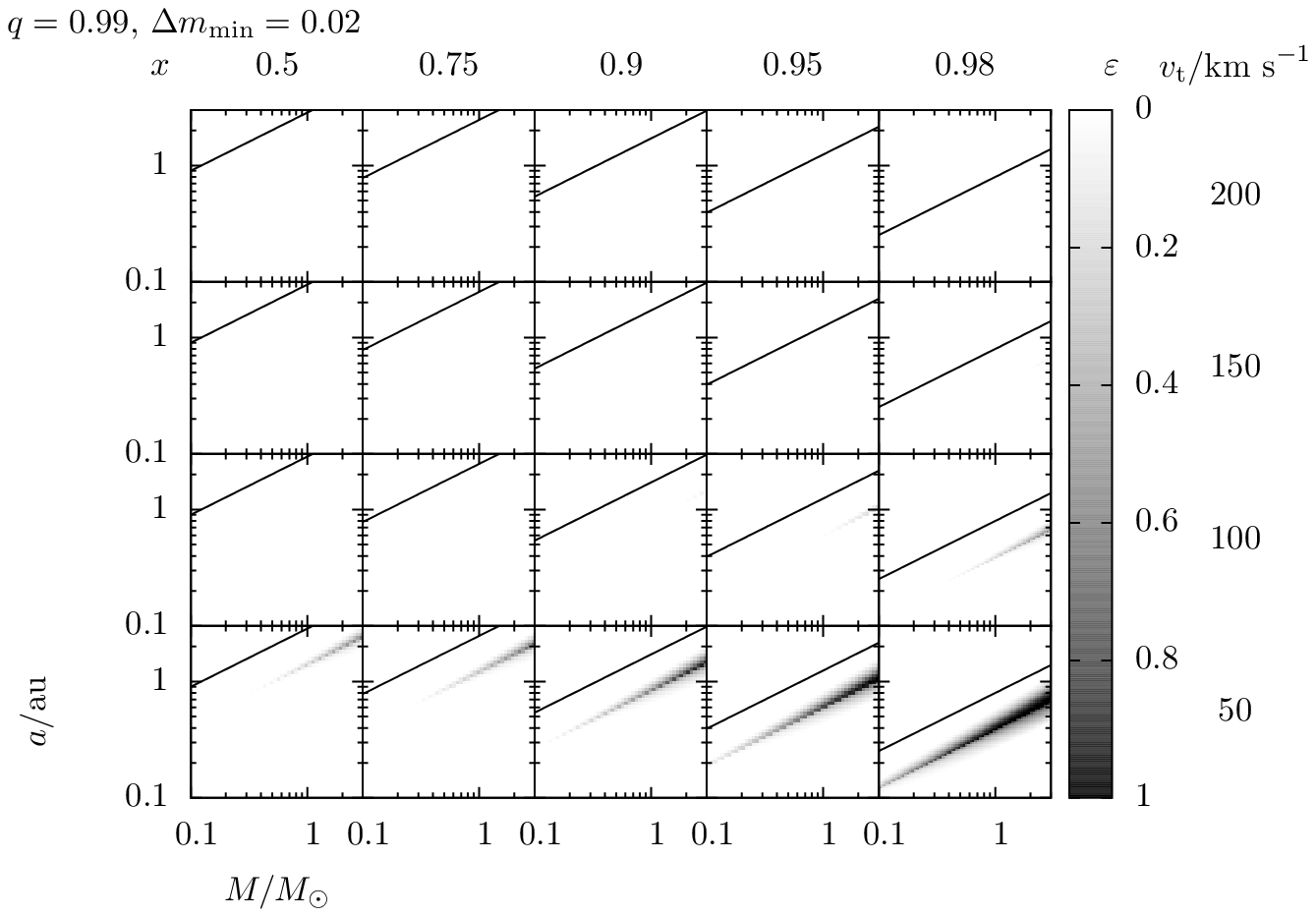}
\hspace{0.02\textwidth}
\includegraphics[width=84mm]{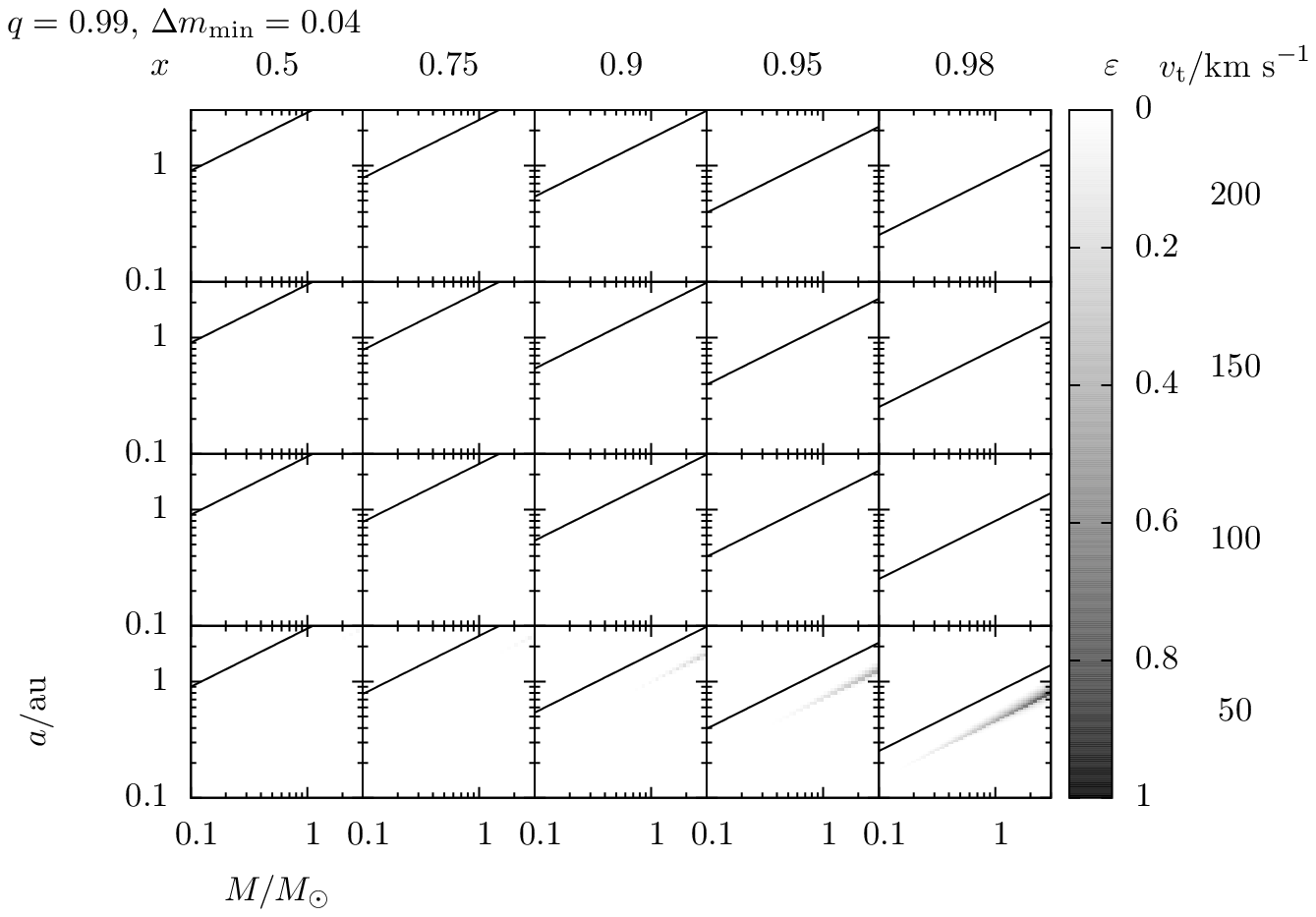}\\
\caption{Maps of detectability $\detectability$ plotted against $a$
  and $M$ for a binary of mass ratio $q=0.99$ and various values of
  the fractional lens distance $x=\dl/\ds$, source velocity $\vt$ and photometric detection threshold $\threshold$. Each small sub-panel is essentially the same as the plot in Figure~\ref{numericalConstraints}, but with different parameter values and a slightly restricted range $0.1\le M/\msun \le 3$, $0.1\le a/\text{au} \le 3$. From left to right, top to bottom panel have increasing values of the photometric threshold $\threshold=0.005,0.01,0.02$ and $0.04$. Moving from left to right, sub-panels have different fractional lens distances $x=0.5,0.75,0.9,0.95$ and $0.98$; the results remain the same under the transformation $x\rightarrow (1-x)$, i.e. there is reflectional symmetry about $x=0.5$. Moving from bottom to top, sub-panels have different source velocity $\vt=50,100,150$ and $200$~\kms. The source distance is fixed at $\ds=8$~kpc. The black line shows the boundary between close and resonant caustic structures, above which we do not plot $\detectability$. As in Figure~\ref{numericalConstraints}, there are points where the calculation of $\detectability$ fails, but
these are not shown for clarity, as they do not impinge on the regions of detectability.}
\label{grid_thresh}
\end{figure*}

Having looked at the role of mass and orbital separation, it is
important to investigate how the detectability of repeating features
depends on other factors. Figure~\ref{grid_thresh} shows detectability
maps similar to that in Figure~\ref{numericalConstraints}, but for a
mass ratio $q=0.99$ and various values of the lens distance, source
velocity and photometric detection threshold. The maps are arranged
into four panels with different photometric threshold values of $\threshold =
0.005, 0.01, 0.02$ and $0.04$ from left to right, top to bottom. Each panel is made up of
a grid of $a$-$M$ maps showing detectability for different fractional lens
distances $x$ in each column, and different source velocities
in each row. Looking first at the grid with $\threshold =
0.01$ (top right), it is clear that the source velocity has a large effect on the
detectability, with  large regions of detectability for $\vt=50$~\kms{}
at all lens positions, which are reduced drastically for
$\vt=100$~\kms. Once $\vt=150$~\kms{} there is only a tiny region of low
detectability for lenses very close to the source (or to the observer,
as $x(1-x)$ is symmetric about $x=0.5$). For $\vt=200$~\kms{} there is
only detectability in the most favourable cases of very high photometric
accuracy and fractional lens distance. This
strong dependence on $\vt$ occurs because the number of orbits
completed by the lens decreases as $\vt$ increases (the
$\smash{\vt^{-2/5}}$ term in Equation~\ref{upperLimit}) but
does not affect the strength of binary features
(Equation~\ref{lowerLimit} is independent of $\vt$) -- there is no
cancellation of the $\vt$ term in the ratio of upper to lower limits,
while there is a partial cancellation for all the other physical
parameters. Unfortunately, the microlensing event rate peaks at
$\vt\sim 200$~\kms, but there is a significant fraction of events with
$\vt<100$~\kms~\citep[e.g.][]{Dominik:2006spd}.

The lens distance does not affect the size of the detectable region as
strongly as the source velocity does, as the upper and lower limits of
the detectable region scale with $x(1-x)$ as similar power laws
($-0.4$ and $-0.5$ respectively). However, this similar scaling means
that the detectable regions move as $x$ changes, occurring at lower $a$,
and increasing in size slightly, as the product $x(1-x)$
decreases. For microlensing events towards the Galactic Bulge, the
event rate peaks at $x\sim 0.8$~\citep[e.g][]{Dominik:2006spd},
whereas for self lensing in the Magellanic clouds $x$ will be close to
unity, $x\approx 0.98$.

The photometric precision of the observations strongly affects the
detectability of repeating features. For $\threshold=0.005$ and $0.01$
we see large regions of detectability for small source velocities, and
for $\threshold=0.005$ even some detectability when
$\vt=200$~\kms{}. As $\threshold$ increases to $0.02$, the detectable
regions shrink significantly and all but disappeared for
$\vt\ge 100$~\kms. For $\threshold=0.04$ there is virtually no
detectability, with only a small chance of detection for the smallest
velocities and largest lens distances. Increasing the threshold
effectively increases the lower limit of $a$ at which binary features
are detectable, while leaving the upper limit unchanged, and like the
source velocity, the photometric threshold has a large effect on the
size of the detectability region. It should be noted that the
detection threshold $\threshold$ is in fact a combination of the effects of
photometric precision and the blending by unrelated starlight, which acts to add a noise component to the measurement of the magnification
caused by the lens. The effect of blending is discussed further in Appendix~\ref{blending}.

Even in the most favourable case of low photometric threshold and source velocity, and high fractional lens distance, the region of detectability does not reach the boundary between close and resonant caustic topologies. This is because as the projected separation increases, near the topology boundary, the secondary caustics move inwards rapidly before merging with the central caustic, decreasing the range over which binary features are detectable. At the same time the orbital period will increase rapidly as the semimajor axis increases. These combined effects mean that in order to see repeating features from a lens with resonant topology, an extremely low source velocity is necessary to allow the lens to orbit in the time the source spends near the resonant caustic.

\begin{figure*}
\includegraphics[width=84mm]{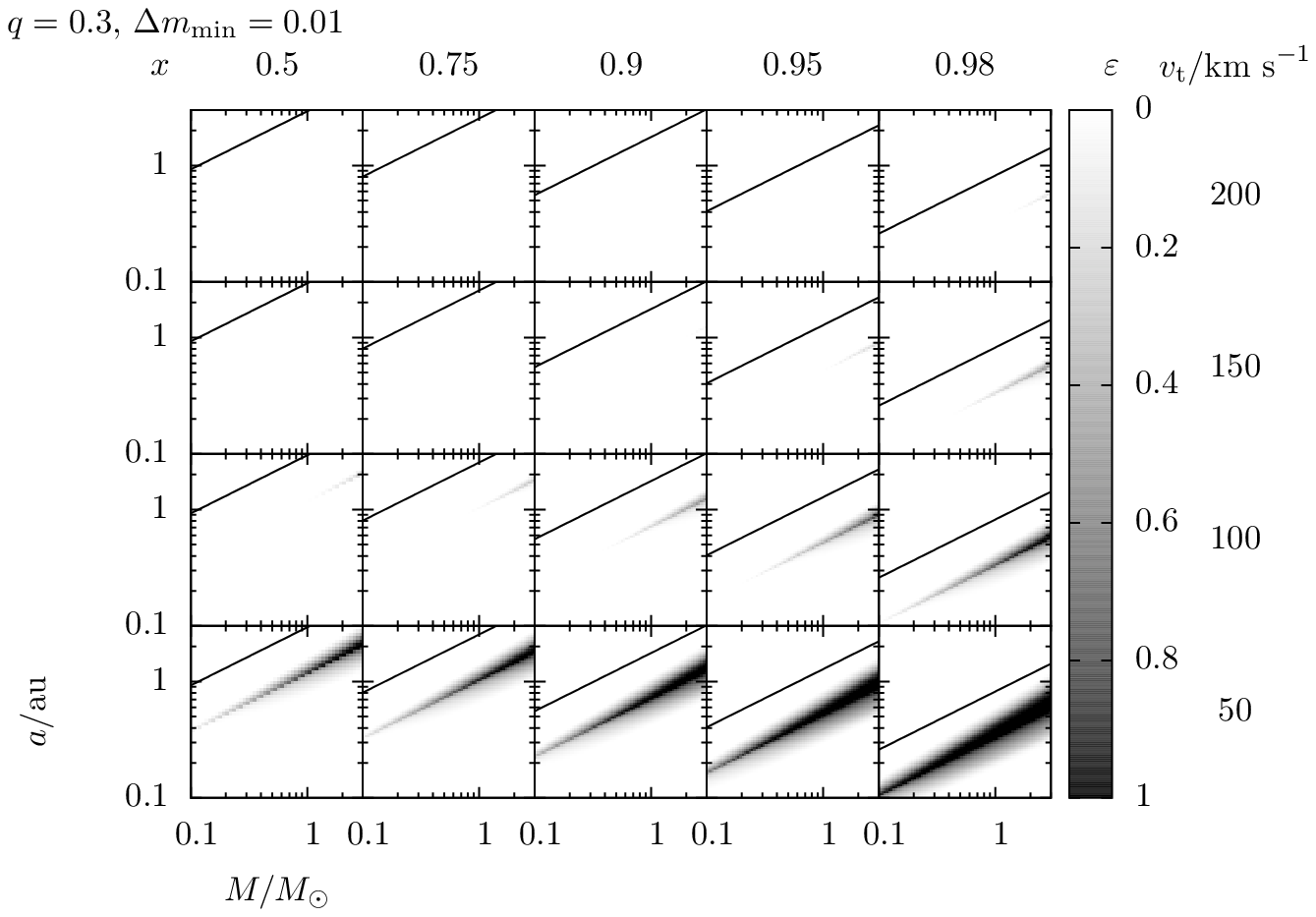}
\hspace{0.02\textwidth}
\includegraphics[width=84mm]{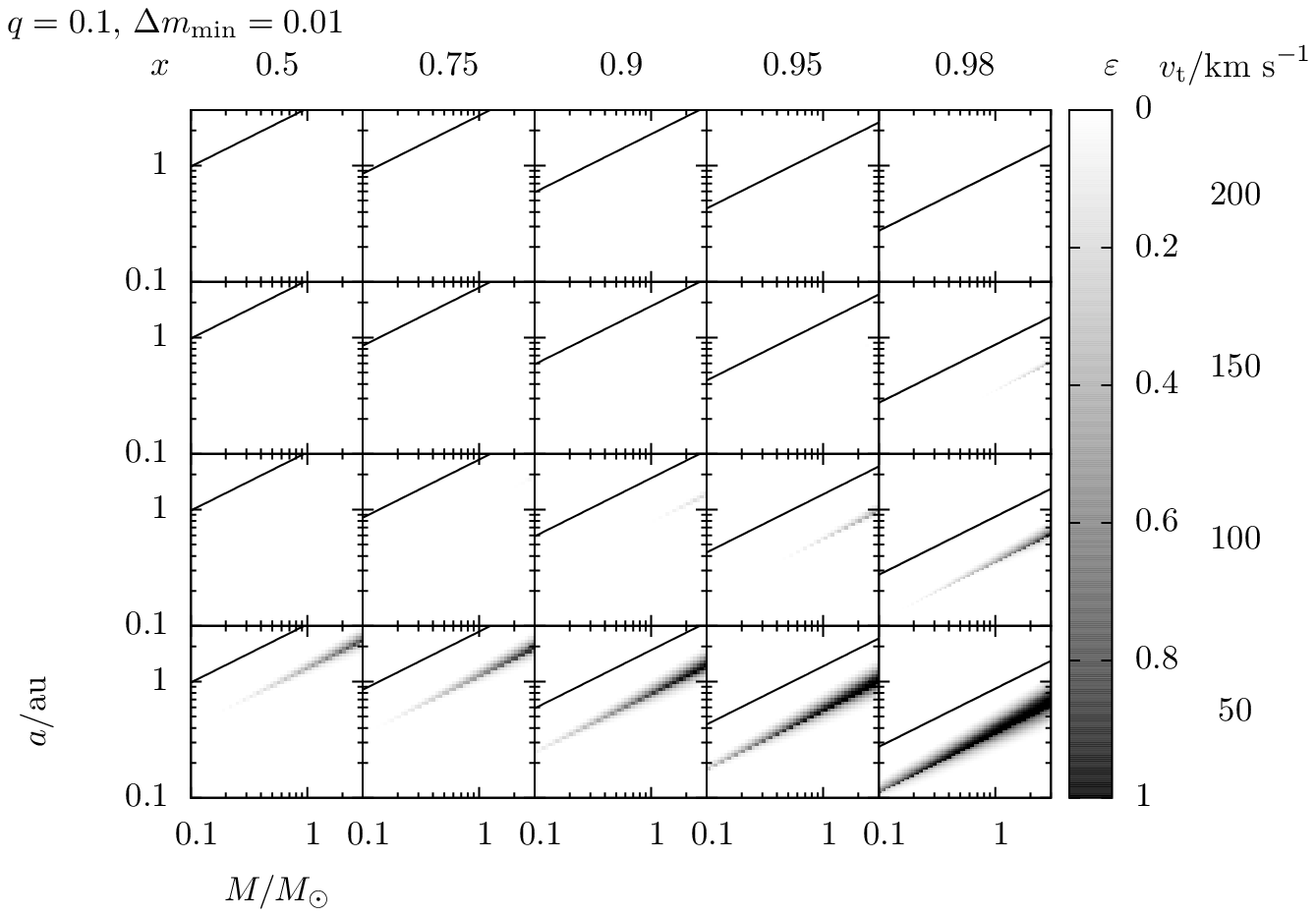}\\
\vspace{12pt}
\includegraphics[width=84mm]{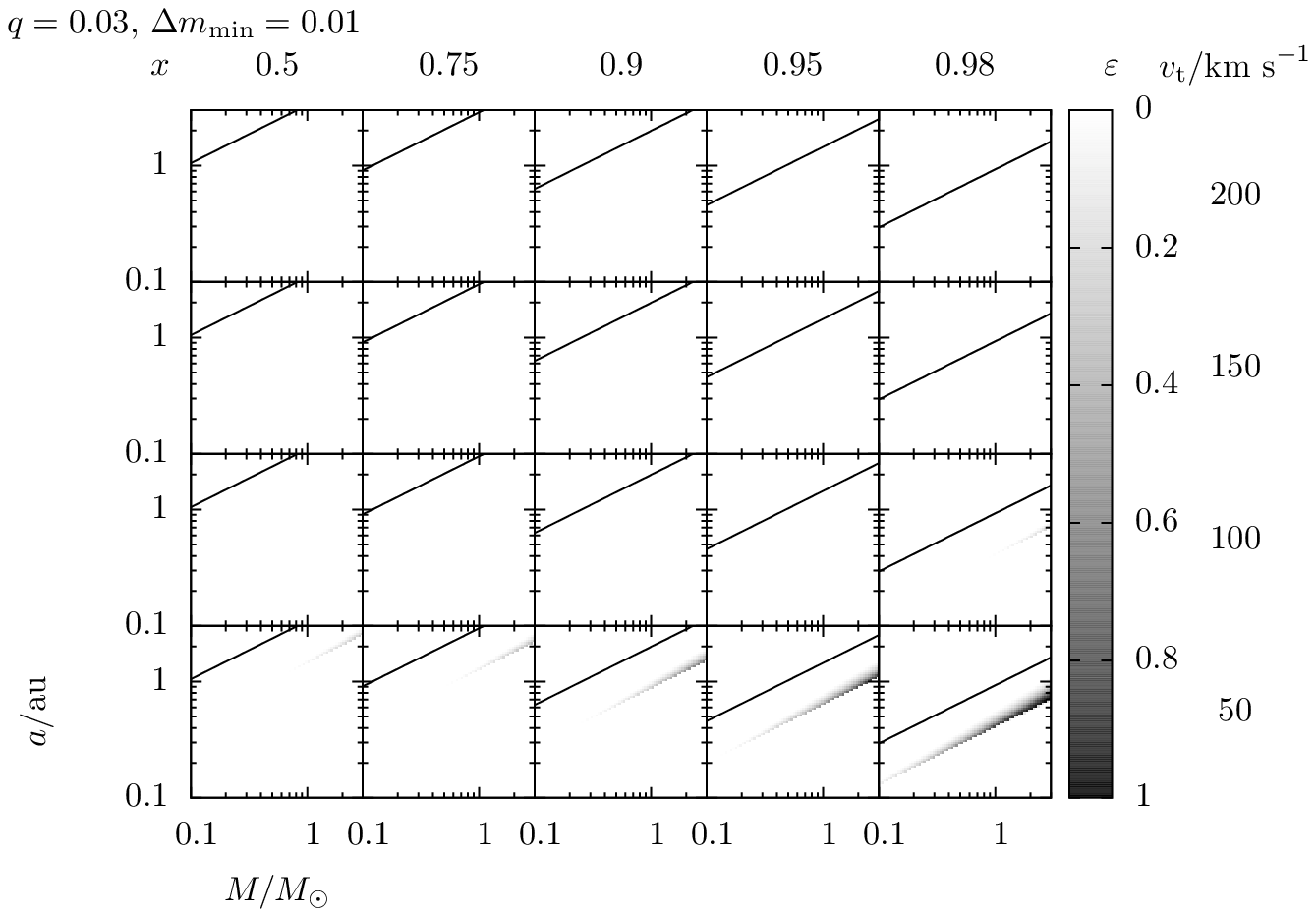}
\hspace{0.02\textwidth}
\includegraphics[width=84mm]{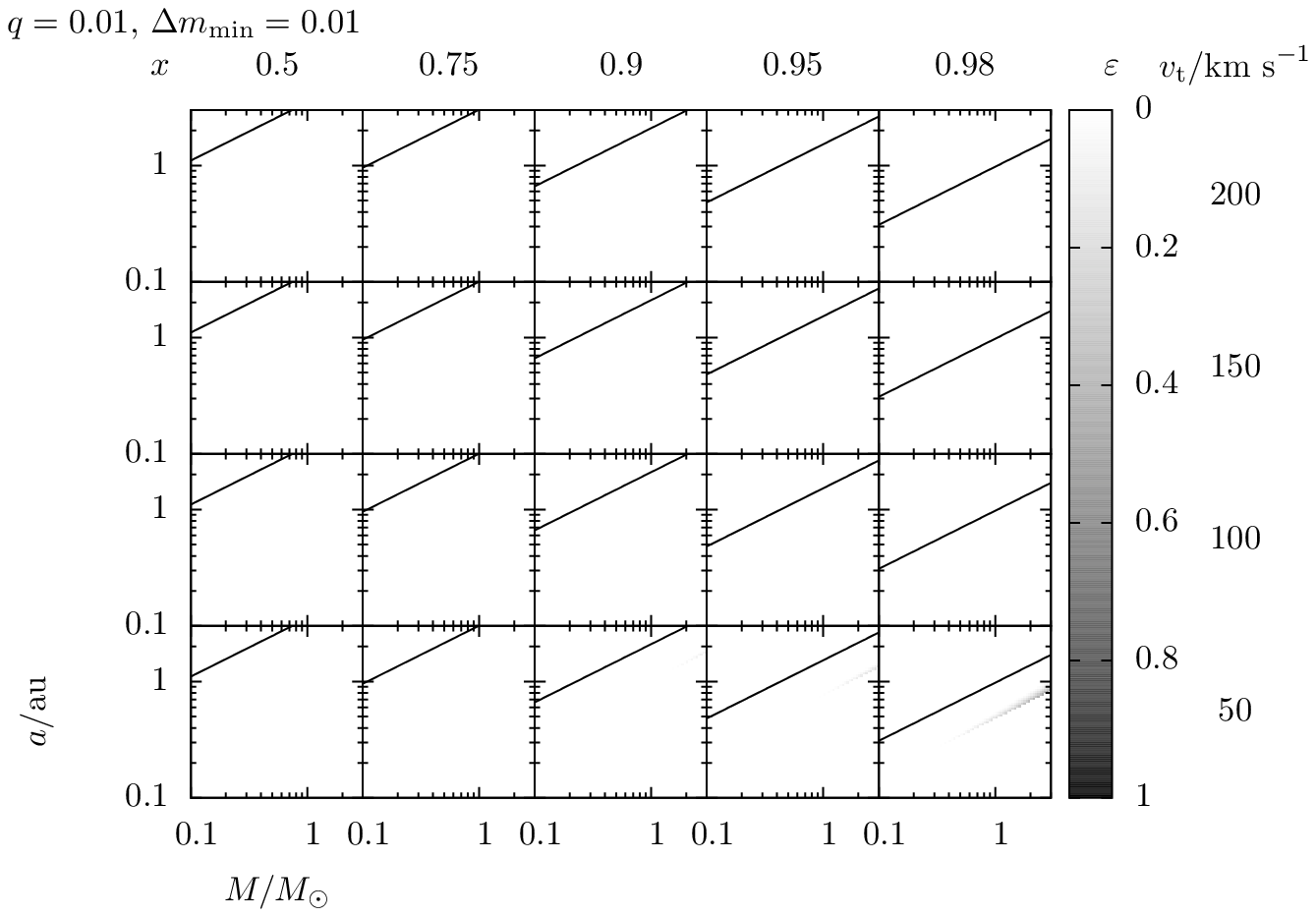}\\
\caption{As Figure~\ref{grid_thresh}, but for differing mass
  ratios. Moving from top left to bottom right the detectability is
  plotted for $q=0.3,0.1,0.03$ and $0.01$. The total mass corresponding to a secondary below the deuterium burning limit,
  $M_{\mathrm{D}}\approx 13 M_{\mathrm{Jupiter}}$, is $M<0.054\msun$,
  $M<0.14\msun$, $M<0.43\msun$ and $M<1.25\msun$ respectively for each
  value of $q$. The photometric detection threshold in each case is $\threshold=0.01$.}
\label{grid_q}
\end{figure*}

Figure~\ref{grid_q} shows the same maps as Figure~\ref{grid_thresh}
but for differing $q$, and the threshold fixed at
$\threshold=0.01$. The maps for $q=0.3$ are similar to those for $q=0.99$, and there is little difference in the size of the region of
detectability. However, once $q$ has fallen to $0.1$, the size of the
detectable region has begun to shrink, such that for higher values of
$\threshold$ (not shown) there is only a very small chance of
detection with small source velocities. For lower mass ratios still,
there are only very small regions of detectability for $q=0.03$ and
effectively zero detectability for $q=0.01$. If we take the boundary
between brown dwarfs and planets to be at $13\mjup$, there is a very
small region of detectability where the secondary lens is a planet,
but the point of the detectable region where the upper and lower
limits meet occurs close to this point, regardless of the mass
ratio. So, there is little chance of detecting repeating features from
a planetary system, unless the photometry is very accurate, the lens
very close to the source, or the source velocity is significantly
smaller than $50$~\kms. Such low velocity events are rare, but are
known to occur, e.g. the event OGLE-1999-BLG-19 had a source velocity
$\vt=12.5\pm1.1$~\kms~\citep{Smith:2002mpp}.

\subsection{How many RRL events will we detect?}

To estimate the RRL event rate we conducted a simulation of a
space-based $H$-band microlensing survey, such as
  Euclid~\citep{Beaulieu:2010eph} or WFIRST~\citep{Bennett:2011wps},
  and a ground-based $I$-band survey, based on
  OGLE-III~\citep{Udalski:1997ots, Udalski:2003ews}. Using the
\besancon population synthesis model of the
Galaxy~\citep{Robin:2003bgm}, including a three dimensional extinction
model~\citep{Marshall:2006ged}, we produced a catalogue of
possible microlensing events following the recipe of
\citet{Kerins:2009smm}. Source stars with magnitudes
$H_{\mathrm{s}}<25$ and $I_{\mathrm{s}}<21$ are drawn from the
\besancon model and lensed by stars of any magnitude in the space- and
ground-based simulations respectively. The lens mass is split up into
two components with a mass ratio $q$ distributed logarithmically in
the range $0.1 \le q < 1$, and orbit with a semimajor axis $a$
distributed logarithmically in the range $0.1 \le a/\text{au} <
4$. Each event is assigned a weighting $w=2\re\vt\uzeromax$
proportional to its event rate, where $\uzeromax$ is the maximum
impact parameter that the event could have and its peak single lens
magnification remain detected at $5\sigma$ above baseline, taking into
account blending. Each event was assigned a blending fraction
$\blendfs^{\prime} \le 1$ drawn from the blending distributions of
\citet{Smith:2007mlb}, with source density $131$ stars per square
arcmin, and seeing $0.7$~arcsec and $1.05$~arcsec for the space-based
and ground-based simulations respectively. This will overestimate the
blending effect for the space-based simulation, as the diffraction
limited PSF for a $1$~m telescope will have a full width half maximum
$\sim 0.4$~arcsec, but \citet{Smith:2007mlb} do not simulate seeing
better than $0.7$~arcsec. The final blending suffered
by the source $\blendfs$ also includes flux from the lens, which is
obtained from the \besancon model assuming it is a single star. The
severity of blending is thus overestimated, especially for the
space-based survey, and as blending has a large effect on the detectability (see Appendix~\ref{blending}), the event rates we estimate will be conservative. However, we do not include the effect of orbital inclination, which can decrease the amplitude of lightcurve features slightly (see Appendix~\ref{inclinationAndEccentricity}), so this optimistic assumption will likely balance the pessimistic blending we apply. The photometric detection threshold was calculated, based approximately on the proposed design of the Euclid mission~\citep{euclid:2009pdd} for the space-based survey, and the OGLE-III setup~\citep{Udalski:1997ots} for the ground-based setup. Total event rates are normalized to rates $\Gamma_{\mathrm{\mu L}} = 7000$~yr$^{-1}$ for the space-based survey~\citep[e.g.][]{Bennett:2002ssm}, and $\Gamma_{\mathrm{\mu L}} = 600$~yr$^{-1}$ for the ground based survey, corresponding roughly with the OGLE-III survey. The rate of RRL events $\Gamma_{\mathrm{RRL}}$ is taken to be 
\begin{equation}
\Gamma_{\mathrm{RRL}} = \frac{\Gamma_{\mathrm{\mu L}}}{W}\sum_i w_i \detectability_i,
\label{rateSum}
\end{equation}
the normalized sum of the product of $\detectability_i$ and $w_i$, the detectability and weight of event $i$ respectively, over all microlensing events, where $W=\sum{}{}w_i$ again summed over all events.

The simulations do not account for the observing strategy, and assume
that frequent monitoring (a few data points per night or greater) is
conducted for a significant fraction of the year (6 months or
greater). It is difficult to assess the impact of seasonal
observability on the probability of detecting repeating features,
without performing detailed detection efficiency simulations. To
account for this we introduce a factor $f_{\mathrm{seas}}$, the fraction
of a year spent continuously observing, which is approximately the probability that an individual feature is `caught'. We must also account for the fact that not every lens is binary. \citet{Raghavan:2010bsf} find that 44 percent of stellar systems are multiple, with mass ratios $q>0.1$, and of these about 20 percent lie in the appropriate semi-major axis range, so we adopt a binary fraction $f_{\mathrm{b}} \approx 0.1$.

For our entire sample of space-based survey events we find that RRL events make up a fraction $(1.1 \pm 0.2)\times 10^{-3}$ of the total microlensing event rate, which corresponds to an event rate $\Gamma_{\mathrm{RRL}} = (7.8 \pm 1.5)f_{\mathrm{seas}}f_{\mathrm{b}}$~yr$^{-1}$. Similarly for the ground-based survey we find that a fraction $(0.5 \pm 0.1)\times 10^{-3}$ of the total microlensing event rate is made up of RRLs, which corresponds to an event rate $\Gamma_{\mathrm{RRL}} = (0.32 \pm 0.06)f_{\mathrm{seas}}f_{\mathrm{b}}$~yr$^{-1}$. In all cases the errors are statistical.

\begin{figure}
\includegraphics[width=84mm]{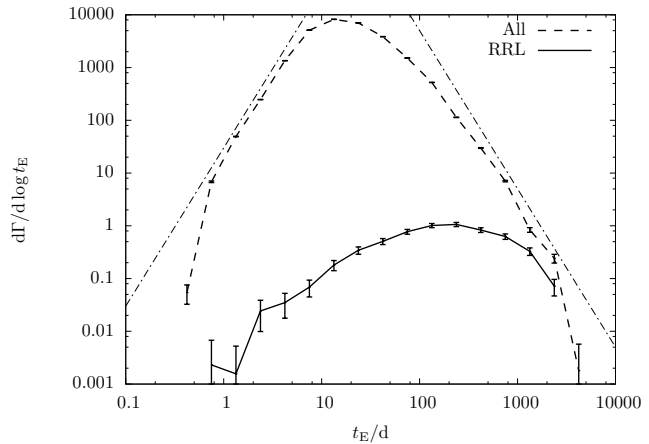}
\caption{Microlensing timescale distributions for detectable RRL
  events (solid line) and all microlensing events (dashed line) for
  the space-based survey. Other than the overall normalization, the
distribution for a ground based survey is very similar. The dot-dashed
lines show the expected asymptotic slope of the timescale
distribution, with power law indices $\pm 3$~\citep{Mao:1996mm}}
\label{timescaleDistribution}
\end{figure}

Figure~\ref{timescaleDistribution} shows the distribution of
microlensing timescales for the detectable RRL events and all
microlensing events in the space-based simulation. The results are
very similar for the ground-based survey, other than the overall
normalization. The distributions do not take into account any
timescale dependence on detection efficiency, or the binary
fraction. The timescale distribution for RRLs shows a peak at
$\tein\sim 200$~d that is at timescales a factor of ten longer than
the overall microlensing timescale distribution. Even at this
timescale however, detectable RRL events make up less than one percent
of the whole. As the timescale increases, the fraction of RRL events
increases. Long timescale events are intrinsically rare, but RRL
events make up a significant fraction of all events with these
timescales, and so such events are good targets to search for RRL
signals. Additionally, their long timescales mean that each event is
observable for many years, and it is possible to obtain dense coverage
of the lightcurve with standard survey-mode observations. The
  timescale distribution for all events agrees well with the expected
  asymptotic behaviour~\citep{Mao:1996mm}, except for the points at
  very small and large $\tein$, where small number statistics are in effect.

Various microlensing surveys have targeted the Galactic bulge, pretty much continuously for roughly twenty years. These survey-mode observations take place over much of the year, so the seasonal observability factor $f_{\mathrm{seas}}$ will be close to unity. There is therefore a good chance that there is of the order of one RRL event in current microlensing data sets. New ground based microlensing surveys, already in operation and due to start in the near future will increase the overall microlensing event rate significantly, and so there is also a reasonable chance of detecting of the order of one RRL over a timescale $\sim 5$ years.

A space-based microlensing survey is proposed for two space missions
which would launch at the end of the decade: ESA's
Euclid~\citep{Beaulieu:2010eph} and NASA's
WFIRST~\citep{Bennett:2011wps}. Such a mission may only spend $2$--$3$ months per year performing a microlensing survey, as the majority of observing time would be spent on dark energy surveys. As such the seasonal observability $f_{\mathrm{seas}}\sim 0.2$ factor would be low, as a high degree of coincidence would be necessary for multiple RRL features to fall within the observing windows. The number of space based RRL detections is therefore likely to be low in reality. However, a dedicated space-based microlensing survey, possibly as a mission extension to Euclid or WFIRST, observing the bulge continuously for most of the year would be very likely to detect RRL events.

\section{Physical parameters from RRLs}
\label{physicalParameters}

The lightcurve of a static binary microlensing event contains information on
the lens, which can be found by fitting the lightcurve with a static
binary microlensing model. Similarly, the lightcurve of a rapidly
rotating lens contains information about the lens and its orbit. In
this section we investigate the information it is possible to extract
from RRL lightcurves, and how this can be done. 

The static binary lens lightcurve for a point source, can be described
with a minimum of seven parameters: three to describe the source
trajectory, usually an impact parameter $\uzero$ and angle $\alpha$,
and the time of closest approach to the origin $\tzero$; one for the
lightcurve baseline $\mbase$; two to describe the lens, the mass ratio
$q$ and projected separation $s$ in units of Einstein radii; and
finally the Einstein radius crossing time $\tein$. The coordinate
system is usually chosen so that both lenses lie on the $x$-axis, and
the origin is the centre of mass; we shall refer to it as the \textit{static
centre of mass system}.  

\begin{figure}
\includegraphics[width=84mm]{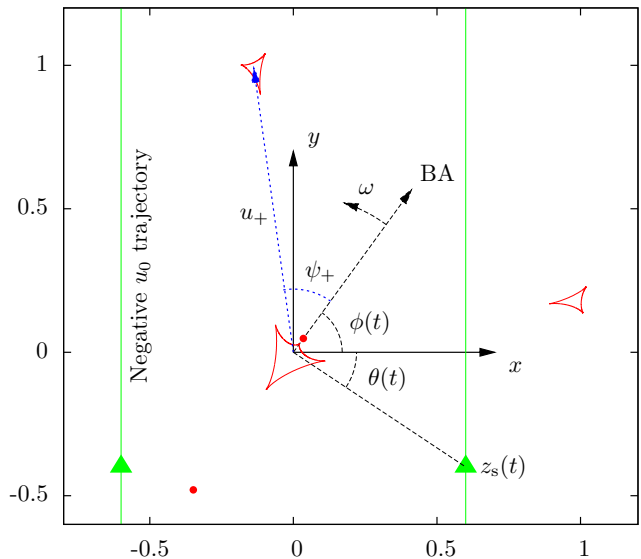}
\caption{Parametrization of the rapidly rotating lenses. Caustics are
  shown as solid red lines, the lens positions as red circles, with the
  primary lens in the positive quadrant, and the source trajectory as
  a solid green line at positive $x$. The green line at negative $x$ shows the trajectory of a source with negative $\uzero$ (see text for more details). The binary axis (BA), which subtends an angle $\phi(t)$
  relative to the fixed $x$-axis, rotates at a frequency
  $\omega=2\upi/T$. ($u_{+}$,$\psi_{+}$) is the position of the
  secondary caustic in polar coordinates fixed to the binary axis. The
  blue dotted line shows the \citet{Bozza:2000cmc} approximation to the
  position of the centre of the secondary caustic
  (Equation~\ref{causticPosition}) for this lens. The lens has the
  parameters $s=0.65$ and $q=0.1$, and lengths are in units of
  Einstein radii.}
\label{parametrization}
\end{figure}

The simplest RRL, with a face-on, circular orbit requires just one
additional parameter, the orbital period $T$, for a total of eight
parameters. In contrast, a full Keplerian orbit requires five
additional parameters (including the period), bringing the total to
thirteen parameters, many of which will be hard to constrain. We
demonstrate below that the eight parameters of the face-on, circular
model can be well constrained by the lightcurve, and parameters can
effectively be `read off' the lightcurve with only a small amount of
algebraic manipulation. It should be possible to use these parameter
estimates in a more detailed modelling analysis, either using the
face-on, circular model (which will be well constrained, should the
face-on, circular orbit approximation apply), or as partial
constraints for a full Keplerian model. This analysis, which we
describe briefly later, can significantly reduce the range of
parameters it is necessary to search in order to find a suitable event
model. In Appendix~\ref{inclinationAndEccentricity} we briefly discuss
the effects of orbital inclination and eccentricity on the lightcurves
and detectability of RRLs, and in Appendix~\ref{parallax} we discuss
the effect of parallax on an RRL lightcurve.

We choose a coordinate
system fixed with respect to the sky, with its origin the lens centre
of mass. As such, the lens components are not fixed. For convenience,
we recast the angle $\alpha \rightarrow \phizero$, where $\phizero$ is
the angle subtended by the primary mass relative to the $x$-axis at
time $\tzero$, and we fix the angle of the source trajectory, such
that the source travels parallel to the $y$-axis. At time $t$
the source is at the (complex) position
\begin{equation}
\zs(t) = \left( \uzero, \frac{t-\tzero}{\tein}\right),
\label{sourcePosition}
\end{equation}
and subtends the angle
\begin{equation}
\theta(t) = \arctan\left( \frac{t-\tzero}{\uzero\tein} \right)
\label{sourceAngle}
\end{equation}
with respect to the $x$-axis. Similarly, the binary axis, which we
define as the line extending from the centre of mass through the
primary mass, subtends an angle
\begin{equation}
\phi(t) = \frac{2\upi}{T}(t-\tzero) + \phizero
\label{phi}
\end{equation}
with respect to the $x$-axis. This parametrization is shown in
Figure~\ref{parametrization}. The parametrization differs from that
recently proposed by \citet{Skowron:2011kos} for orbiting binary
lenses, which is best suited for binaries with orbits much longer than
the microlensing timescale. The \citet{Skowron:2011kos}
parametrization is expressed in terms of the $3$-dimensional position
and velocity of one lens component, as the ``on-sky'' position components
will be well constrained, the ``on-sky'' velocity components may be
well constrained and the radial position and velocity are likely to be
poorly or not constrained. However, as we will show, for an RRL it is
the orbital period and phase angles that will be well constrained, so
it is better to couch the problem in terms of these quantities.

Many of the features in a close binary lens magnification map are
radial, or approximately so. This makes them ideal for measuring the
rotation rate of the lens. A feature occurs on the lightcurve when a
magnification pattern feature sweeps over the source. A radial feature
that subtends the angle $\psif$ relative to the binary axis will occur
on the lightcurve when
\begin{equation}
\theta(t) = \phi(t) + \psif.
\label{generalFeature}
\end{equation}
By solving this equation we can use the timing of repeated
features to easily obtain approximate measurements of some of the
lens parameters. This means that many of the lens parameters can be
`read-off' the lightcurve, and it is possible to build an approximate
model of the lens quickly, without complex analysis. For such
estimations, the most important magnification map features are the
magnification arms (shown in Figure~\ref{exampleMagMap}, which extend
from the central caustic to the secondary caustics), and a planetary
demagnification, a feature that is only present for lenses with small
mass ratios, which is a region of demagnification relative to the
single lens that lies between the secondary caustics, with its minimum
lying along the binary axis. Both features are complementary, as in
equal mass ratio binaries the planetary demagnification does not
occur, but the magnification arms are strong and very close to radial,
while in low mass ratio binaries, the magnification arms are weaker
and less radial, but the demagnification region is strong.

\begin{figure}
\includegraphics[width=84mm]{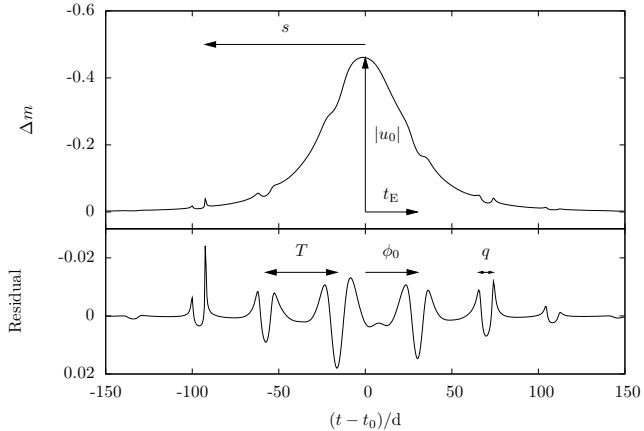}
\caption{An example lightcurve by an RRL showing how lightcurve
  features relate to the parameters of the lens. The lens has
  parameters $\tein=30$~d, $T=38$~d, $s=0.3$, $q=0.1$, $\uzero=0.8$, $\phizero=2.14$}
\label{parametersLightcurve}
\end{figure}

Figure~\ref{parametersLightcurve} shows a lightcurve where features repeat
strongly five times. The first step to estimating RRL parameters is to
fit the lightcurve with a \paczynski curve. This is relatively
trivial, and most RRL lightcurves will approximate a \paczynski curve
with only small deviations. This fit allows an accurate estimation of
the parameters $\tzero$, $\tein$ and $|\uzero|$, the last down to an
ambiguity in sign, which corresponds to the source passing the lens centre on its left (positive $\uzero$) or right (negative $\uzero$) and imposing that the lens always rotates anti-clockwise. This \paczynski model completely describes the
source trajectory, and hence defines the left hand side of
equation~\ref{generalFeature}. The orbital period can now be estimated
by timing two occurrences of the same magnification pattern
feature. The period is not simply the time elapsed between two
features, because the source moves during this time. Instead, by solving equation~\ref{generalFeature} we can find the relation
between the period $T$, and the time of two consecutive occurrences of the same
magnification pattern feature at times $t_1$ and $t_2$
\begin{equation}
T = \frac{2\upi}{2\upi + \left[\theta(t_2)-\theta(t_1)\right]} \left(t_2-t_1\right),
\label{periodAccurate}
\end{equation}
where the fraction is the number of orbits the source completes between the two source encounters. The degeneracy in the measurement of the sign of $\uzero$ affects this equation, due to the presence of the $\theta(t)$ terms, but can be
resolved if more than one pair of features is available for estimating
$T$, as only one value of $\uzero$ will give consistent estimates of
$T$ for different feature pairs.

With an estimate of the period, if we know the angle subtended by a
feature on the magnification map $\psif$, we can also estimate the
phase angle $\phizero$, again taking into account the source motion
\begin{equation}
\phi_0 = \theta(\tf)-\psif-\frac{2\upi}{T}(\tf-\tzero).
\label{phi0first}
\end{equation}
The planetary demagnification region has $\psif=0$, which makes this
task simple. However, the demagnification may not be obvious, or if
the mass ratio of the lens is high, may not be present. In these cases
it is necessary to know $\psif$ for the magnification arms. Knowing
that they extend from the central caustic (roughly at the centre of
mass) to the secondary caustics, we need only know the position of the
secondary caustics to estimate $\psif$. \citet{Bozza:2000cmc} has
derived analytical approximations to the position and shape of
secondary caustics in close lenses using a series expansion of the
Jacobian, critical curves and caustics for $s\ll 1$. He finds that the
secondary caustics are located at
\begin{equation}
z_{\pm} \simeq \frac{1}{s(1+q)}\left[ \begin{array}{cc}
    (1-q)(1-s^2)\\\pm\sqrt{q}(2-s^2) \end{array}  \right],
\label{causticPosition}
\end{equation}
in the static centre of mass system. Figure~\ref{parametrization}
shows that this expression is reasonable even when $s$ is quite
large. If we assume the magnification arms are radial, we can use Equation~\ref{causticPosition} to approximate the angle of the magnification arms, to second order in $s$, as 
\begin{equation}
\psi_{\pm} \simeq \arctan \left[ \frac{\pm\sqrt{q}(2+s^2)}{1-q} \right],
\label{psipm}
\end{equation}
which is relatively insensitive to the lens separation $s$. It is useful to note
the asymptotic behaviour: $\psi_{\pm}\simeq \pm 2 q^{1/2}$ as
$q\rightarrow 0$ and $\psi_{\pm} \rightarrow \pm\upi$ as $q \rightarrow
1$. While the dependence of $\psi_{\pm}$ on $q$ implies an ambiguity
in the estimation of $\phizero$, the corollary is that we can estimate
the mass ratio from the timing of features as well. Using the times of
consecutive magnification arm crossings, $t_{+}$ and $t_{-}$, we have
\begin{equation}
|\psi_{\pm}| = \frac{1}{2}\left|\theta(t_{-})-\theta(t_{+}) - \frac{2\upi}{T} \left(t_{-}-t_{+}\right)\right|.
\label{psipmsol}
\end{equation}
This value can be substituted into equation~\ref{phi0first}, and equation~\ref{psipm} can then be solved for $q$.

The remaining parameter that we are interested in is the lens
separation $s$. The angle of features is essentially independent of $s$, so it is
not possible to estimate $s$ by timing features. However, by noting
that the magnification pattern becomes essentially featureless beyond the
secondary caustics (see Figure~\ref{exampleMagMap}), and that the
position of the caustics does depend on $s$, it is possible to
estimate $s$ from the lightcurve. Unfortunately the secondary caustics
are very small, and in most events, they will not pass directly over
the source, so the estimate will not be very accurate. The best
estimate of the position of the caustic will be derived from the
largest peak due to a magnification arm in the wings of the lightcurve
(e.g. the peak at $t\approx -90$~d in
Figure~\ref{parametersLightcurve}). This will occur when the radial
source position approximately coincides with the radial caustic
position, so that $|\zs|^2 \approx |z_{\pm}|^2$. Using
Equation~\ref{causticRadialPosition}, to first order, we can write
\begin{equation}
s \approx \left[\uzero^2 + \left( \frac{t_{\mathrm{c}}-\tzero}{\tein}\right)^2\right]^{-1/2},
\label{destimate}
\end{equation}
where $t_{\mathrm{c}}$ is the time of the peak due to the caustic.

We have outlined how the parameters of an RRL can be estimated from
pairs of feature timings in the case of the simplest RRL. However, in
a given event there may be many repetitions, and better parameter
estimates can be obtained by considering all the lightcurve features
simultaneously. For a given magnification pattern and source
trajectory it is possible to compute a timing model by finding all
possible solutions of equation~\ref{generalFeature}, $\theta(t) =
\phi(t) + \psif$ for each feature. By extracting the occurrence time of
all the lightcurve features it is possible to fit timing models to this
timing data. It is also possible to add additional features to this
timing model, such as the effects of inclination and eccentricity by
modifying the function $\phi(t)$, or microlensing parallax by
modifying $\theta(t)$. This modelling may be significantly faster than a full
lightcurve fitting analysis, especially when additional effects are
included, as there is no need to calculate finite-source
magnifications. While it will not fully remove the need for lightcurve
fitting, it will significantly narrow down the range of parameters
over which lightcurve fitting has to search.

We have shown that it is possible to estimate the parameters of an RRL
lightcurve, but what we would really like is to be able to
measure the physical parameters of the lens, most importantly the lens
mass and the binary separation in physical units. Compared to a static
binary lens, we have one additional piece of information with which to
infer $M$ and $a$: the orbital period. \citet{Dominik:1998mrb} has
shown that by combining the orbital period $T$ and the lens separation
$s$, it is possible to write down a mass-distance relation
\begin{equation}
M = \frac{T^4}{C^6s^6x^3(1-x)^3\ds^3},
\label{massDistance}
\end{equation}
which relates the mass to the lens distance through known quantities,
assuming the source distance is known from its colour and magnitude;
the constant $C=2.85\msun^{-1/2}$~au~kpc$^{-1/2}$ when the
period is measured in years and the source distance in kpc. As
demonstrated in Appendix~\ref{inclinationAndEccentricity} it is likely
that if the orbit is inclined it will be possible to measure the
inclination and account for projection, so that the value of $s$
measured is a good approximation of $a/\re$. This means that as
equation~\ref{massDistance} has a minimum at $x=0.5$, we can place a
firm lower limit on the mass of the lens, and an upper limit on the
semi-major axis. 

\begin{figure}
\includegraphics[width=84mm]{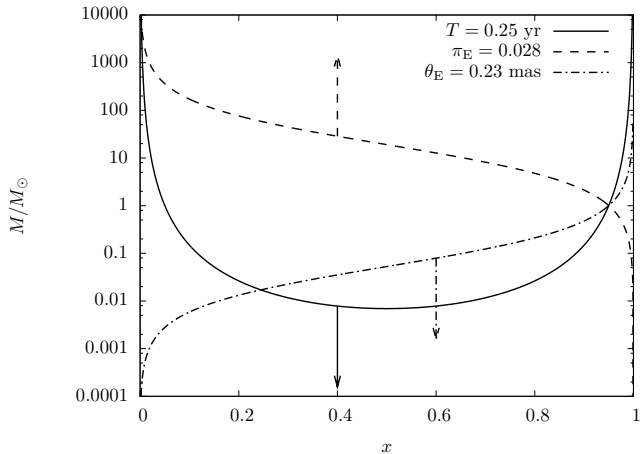}
\caption{Plot of the various mass-distance relations for the event shown in
  Figure~\ref{parametersLightcurve}, labelled by the parameter
  measurement that would allow their definition. The arrows point into the
  region that is \textit{allowed} should only an upper limit on $T$,
  $\pi_{\mathrm{E}}$ or $\thetae$ be available. If the period $T$ is
  measured along with only one of $\pi_{\mathrm{E}}$ or $\thetae$, the
  mass and distance to the lens can not be determined uniquely, but
  even a relatively weak upper limit on the other parameter may be
  sufficient to rule out one possible solution; note however that a
  lack of finite-source effects places a lower limit on $\thetae$.}
\label{Mxrelation}
\end{figure}

To improve on the mass-distance relation, an additional piece
  of information is needed to break the degeneracy. This can be done
  by measuring $\pi_{\mathrm{E}}=\text{au}(1-x)/\re$, the microlensing
  parallax~\citep{Gould:1992mss}, or by measuring $\thetae=\re/\dl$,
  the angular Einstein radius, through detection of finite-source effects~\citep{Gould:1994mte,Nemiroff:1994mfs,Witt:1994fs},
  or direct detection of the lens once it has separated from the source~\citep{Alcock:2001ddl,Bennett:2006ohs,Bennett:2007phc,Kozlowski:2007dd}.
Measurement of either $\pi_{\mathrm{E}}$ or $\thetae$ allow a second mass-distance
relation to be written, for $\pi_{\mathrm{E}}$~\citep{Gould:1992mss}
\begin{equation}
M = \frac{\text{au}^2(1-x)}{C^2 x \ds \pi_{\mathrm{E}}^2},
\label{piM-x}
\end{equation}
or similarly for $\thetae$~\citep{Gould:1994mte,Nemiroff:1994mfs}
\begin{equation}
M = \frac{\thetae^2 x \ds}{C^2 (1-x)},
\label{thetaEM-x}
\end{equation}
if $\thetae$ is measured in mas. One of these relations can then be combined with
equation~\ref{massDistance} to yield two possible solutions to the
mass and distance. This can be seen in Figure~\ref{Mxrelation}, which
plots the mass-distance relations for the event shown in
Figure~\ref{parametersLightcurve}.\footnote{Note that parallax or
  finite-source effects were not included in the model used to plot
  the lightcurve} The $\pi_{\mathrm{E}}$- and
$\thetae$-lines cross the $T$-line in two places: once at the true
parameter values $x=0.95$, $M=1\msun$, and once at other values of $M$
and $x$ which are different for each relation. With a measurement of
only one of $\pi_{\mathrm{E}}$ or $\thetae$, it is not possible to
uniquely determine the mass and the distance. This is likely to be the
case, as finite-source effects are most likely in lenses close to the
source, while parallax is most likely in lenses close to the
observer. However, even a crude limit on the unmeasured parameter may be enough to rule out one possible solution; e.g. an upper limit on $\pie$ from the lack of parallax effects may allow the solution with smaller $x$ to be ruled out, or a lower limit on $\thetae$ from the lack of finite-source effects may allow the solution with larger $x$ to be ruled out. Direct detection of the lens may require a very
long time baseline as RRL
features are most detectable in events with low lens-source proper
motions. However, RRLs are more likely to be more massive than the
average lens, and hence brighter, and the diffraction limit of $30$--$40$~m
class telescopes, such as the Thirty Metre Telescope
(TMT),\footnote{http://www.tmt.org} the Giant Magellan Telescope (GMT),\footnote{http://www.gmto.org} or the European Extremely Large
Telescope
(E-ELT),\footnote{http://www.eso.org/public/teles-instr/e-elt.html} may be
sufficient to resolve the lens and source in a reasonable time.

\section{Discussion and conclusion}
\label{discussion}

Although the phenomena of microlensing by lenses with rapid orbital motion has been discussed in the literature previously~\citep{Dominik:1998mrb,Dubath:2007rrl}, the realities required to estimate its occurrence rate have not properly been accounted for. In this paper, we have established a firm theoretical basis for rapidly rotating lenses, and used it to estimate the range of parameters over which they are detectable and the rate that they are expected to be observed. We find that RRLs with masses and orbital radii typical of binary stars are detectable, and that there is a reasonable chance that they will be detected, either in current microlensing data sets or in ongoing or near-future microlensing surveys.

In calculating these rates we have actually used the relatively stringent criteria of requiring that two or more lightcurve features from the same orbital phase are detected in the lightcurve. If we relax this repetition requirement somewhat, to include lenses that display significant signs of orbital motion (say several degrees rotation per $\tein$), the event rate will increase significantly, as lenses can then have larger orbits and hence stronger lightcurve features. We have shown in a previous paper that orbital motion is detectable in a large fraction ($\sim 15$~percent) of binary lenses with detectable binary lensing features and orbital periods comparable to the microlensing timescale~\citep{Penny:2011omm}. 

We have detailed how the features of an RRL lightcurve can be used to measure its period and potentially measure its mass. Even if features do not repeat, if several features are detectable in the lightcurves of binary lens events, then the techniques we have outlined for timing features and extracting parameter estimates may be of some use in their analysis. Without repeating features, the orbital period may not be constrained as accurately, but it should be possible to place constraints on the lens mass and orbit in many cases.

So far, we have neglected to discuss the prospects for positively
identifying RRL events from other events which may mimic their
features. Periodic features may also be induced by orbital motion in
the observer and source planes, or intrinsic variability in the source
or a blend star. In the observer plane, the period of orbital parallax
effects is well defined, and unless the lens has an orbital period
similar to $1$~yr it is unlikely the effects will be confused. Even if
the orbital period is close to one year, the shape of features in the
lightcurve are likely to be different. Orbital effects in the source
plane may be more difficult to exclude, as the period is not fixed. If
there is only a single luminous source \citep[the xallarap
case,][]{Paczynski:1997, Han:1997bsl, Rahvar:2009mxe}, a timing
analysis similar to the one we proposed for the lens can be performed
for the source. This analysis is somewhat easier and more precise for
xallarap as there are no complicated features in the magnification
pattern. If this analysis is insufficient to separate the two cases
then the shape of lightcurve features should differentiate the two
interpretations. In the case where both sources are luminous, the
lightcurve can take a more complicated shape, which may more closely
resemble that of an RRL~\citep[e.g.][]{Cherepashchuk:1995bsm,
  Han:1997bsl}. Even in this case, timing analysis for maxima and
minima of the lightcurve should be easier than for RRLs, and full
lightcurve modelling starting from timing analysis solutions will
likely be able to differentiate the two scenarios. Finally, source
variability that is not detectable at baseline, but becomes detectable
with the increased photometric accuracy thanks to microlensing
magnification may also produce similar lightcurve features.

It is worth noting that we should naively expect the rate of RRL/significant lens orbital motion events to be similar to the rate of binary source orbital motion events, as the factors that govern their occurrence, such as the ratio of orbital separation to the Einstein ring, and the ratio of orbital to microlensing timescales will have similar distributions in the lens and source populations. Naively, we would expect the rate of parallax events to be roughly ten times greater than the rate of RRL events with orbital periods $\sim 1$~yr, as the binary fraction is $\sim 0.1$ while the observer is always orbiting. It is worth comparing this with the number of reported single lens parallax events, $\sim 20$--$50$~\citep[e.g.][and references therein]{Poindexter:2005mpa, Smith:2005pml}, while $\sim 10$ events have been successfully fit with xallarap models \citep{Smith:2003ap, Poindexter:2005mpa}. In contrast, only one binary lens event has been successfully fit with a binary lens model that shows significant rotation, MACHO-97-BLG-41~\citep{Albrow:2000rbl}, which rotates at a rate $\sim 4\degr$ per $\tein$, a low rate compared to RRLs, that is detected thanks to the source crossing the central and one secondary caustic, as opposed to the smaller, more continuous features of RRLs. It is possible therefore that many events with significant rotational orbital motion signatures have not been modelled, or have been interpreted as xallarap events. It is thus important that any event that is modelled with xallarap also be tested with an orbiting binary lens model.

\section*{Acknowledgements} 
We would like to thank the referee Andy Gould for helpful comments that have improved the paper. MP acknowledges the support of an STFC studentship.

\bibliographystyle{mn2e.bst}
\bibliography{mn-jour,rrl}

\appendix

\section{What is happening to the images?}
\label{images}

The image configuration of a point-mass lens consists of two images: a
major image, of positive parity and magnification $\mu_{+} \ge 1$,
outside the Einstein ring and a minor image of negative parity and
magnification $\mu_{-} < 0$, inside the Einstein
ring~\citep[e.g.][]{Refsdal:1964gl,Liebes:1964gl}. The addition of a
second mass to the lens causes an additional image of negative parity
to be produced if the source does not lie within a
caustic~\citep{Schneider:1986bgl}. If the lens is far from resonance,
i.e. $s\ll 1$ or $s\gg 1$, two of the three images can still be
associated with the major and minor images of the single lens, while
the new third image is labelled a tertiary image.

\begin{figure}
\includegraphics[width=84mm]{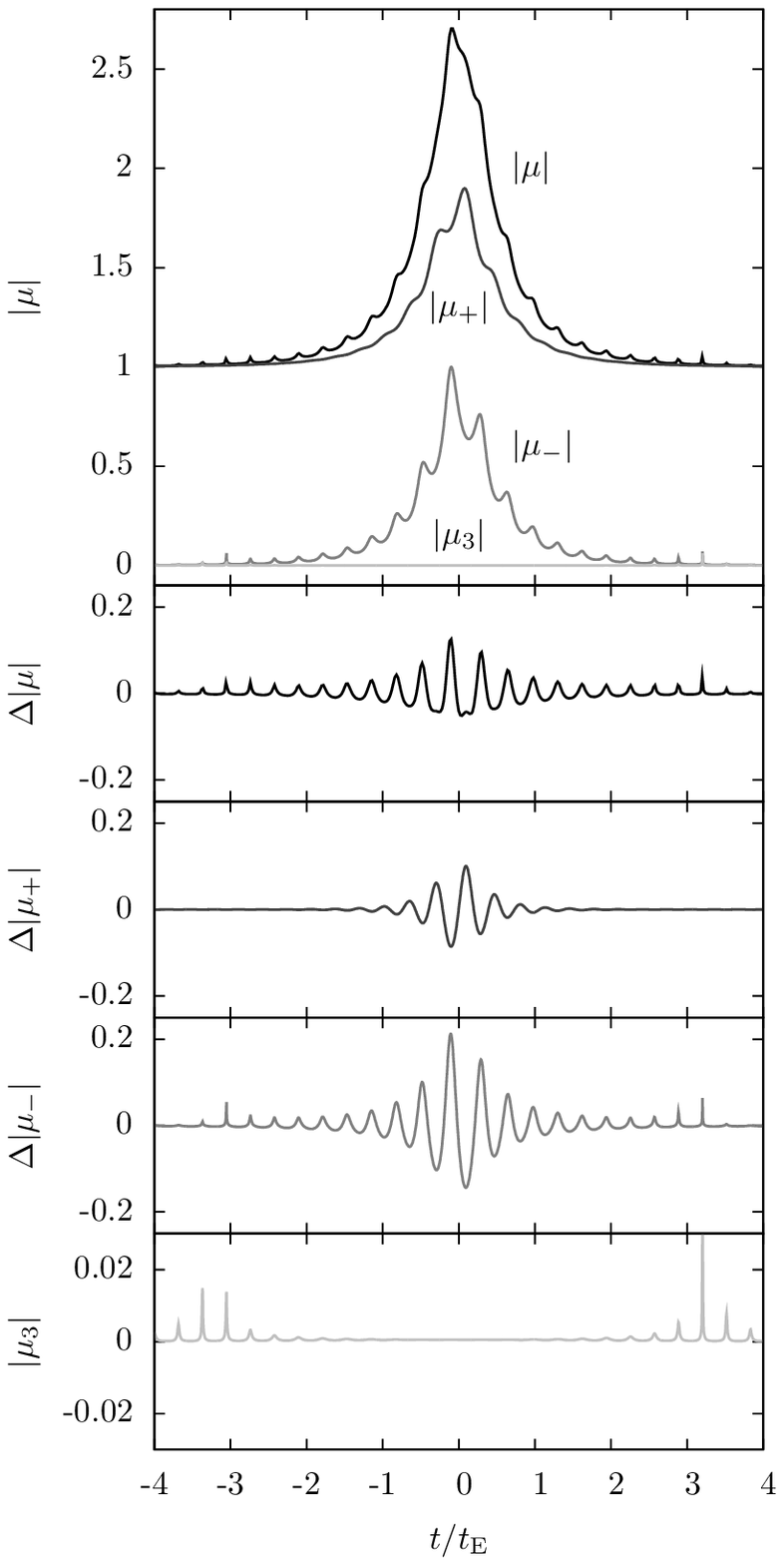}
\caption{Lightcurves and residuals for each image of a microlensing
  event with repeating features. The top panel shows the absolute
  magnification of the combined images ($|\mu|$), and the individual
  major ($|\mu_{+}|$), minor ($|\mu_{-}|$) and tertiary ($|\mu_{3}|$) images in different shades of grey. The central panels show the absolute magnification
  residual with respect to the single lens form for all images
  combined, the major image and the minor image, going from top to
  bottom, respectively; the bottom panel shows the absolute
  magnification of the tertiary image, which has no single lens
  counterpart. The event has the unrealistic parameters $\uzero=0.4$,
  $s=0.3$, $q=1.0$ and $\tein/T=10$.}
\label{whichImages}
\end{figure}

It is interesting to study what is happening to each of the three
images during the course of an RRL event. \citet*{Dubath:2007rrl}
study the effects of an orbiting close binary lens on the major image,
by casting the lensing potential as a time varying quadrupole. They
show that the major image can exhibit significant time dependent
deviations from the single lens form when it is highly magnified, and
go on to calculate the expected rate of events showing such
deviations. Unfortunately, they neglect to treat both the tertiary
image and the minor image, the latter of which will be magnified to a
similar degree, as
$|\mu_{+}|-|\mu_{-}|=1$~\citep[e.g][]{Refsdal:1964gl,Liebes:1964gl}.

In Figure~\ref{whichImages} we plot the lightcurves of
all three images for an unrealistic lens with repeating features. In
the top panel of the figure the combined lightcurve (that which would
be observed) is shown in red, and the absolute magnification of the
major $|\mu_{+}|$, minor $|\mu_{-}|$ and tertiary images $|\mu_{3}|$
is plotted in green, blue and magenta respectively. The second panel
from the top shows the residual of the lightcurve relative to the lightcurve
of a single lens of the same total mass, and the third and fourth
panels show the absolute magnification residuals of the major and
minor images with respect to their single lens counterparts; the
bottom panel shows the absolute magnification of the tertiary image
(note the different scale), which does not have a counterpart for the
single lens. Each image shows a strikingly different pattern of
features: the major image is only significantly perturbed from its
single lens form when the source is within $\sim \re$ of the centre of
mass, while the minor image shows significant perturbations out to the
position of the secondary caustics. It is only when the source is
close to the secondary caustics that the tertiary image is magnified
significantly. However, the most important aspects in relation to the
\citet{Dubath:2007rrl} event rate calculation are that firstly, the
minor image can experience larger perturbations than the primary
image, and secondly that the perturbations of the major and minor
images are out of phase by $\upi$, in such a way that the peak to
trough amplitude of the deviations in the total amplification
lightcurve can be significantly reduced from those of the lightcurves
of individual images.

\section{Additional factors affecting RRL Detectability}
\label{additionalFactors}

In the main text we have mentioned a number of additional effects that
can affect the form of an RRL lightcurve and its detectability. Here
we briefly outline the three most important effects and the impact
they will have on RRL lightcurves and detectability.

\subsection{Blending}
\label{blending}

\begin{figure}
\includegraphics[width=84mm]{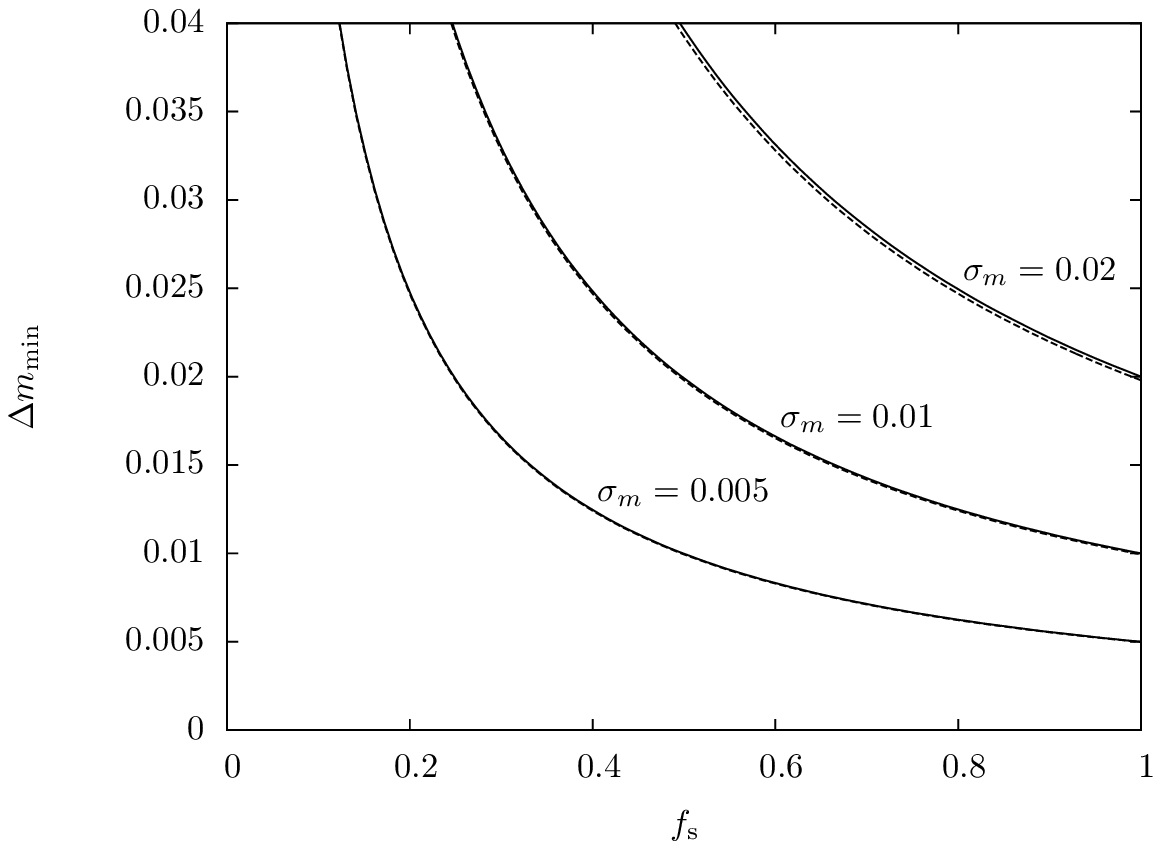}
\caption{The effect of blending on the photometric detection
  threshold. The effective threshold $\threshold$ is plotted against
  the ratio of source to total blend flux $\fblend$ for three values
  of photometric precision $\sigma_m$. The solid lines show the exact
  value, whereas the dashed line shows the approximation for small
  $\sigma_m$.}
\label{blendingEffect}
\end{figure}

For a given photometric precision $\sigma_m$
magnitudes, the effective threshold at the event baseline is 
\begin{equation}
\begin{array}{rl}
\threshold & = 2.5\log\left(10^{0.4 \sigma_m}-1+\fblend\right)-2.5\log
\fblend \\
           & \simeq 2.5 \log \left( 1 + 0.92 \frac{\sigma_m}{\fblend}
\right),\\
\end{array}
\label{blendedThreshold}
\end{equation}
where the approximation applies for small $\sigma_m$ and $\fblend$ is
the fraction of the total light at baseline contributed by the unlensed
source. Figure~\ref{blendingEffect} shows this for various values of
the photometric threshold. It is clear that only with the most
accurate photometry will it be possible to detect RRL features when
the blend contributes most of the flux, and for less accurate
photometry, $\sigma_m\approx 0.02$ even a small amount of blending
will significantly affect the detectability of features. The effect of
blending decreases as the magnification increases, but we wish to see
features over the entire lightcurve, and only a small region of the
lightcurve will be magnified enough to significantly reduce the effect
of blending.

\subsection{Finite-source effects}
\label{finiteSource}

\begin{figure}
\includegraphics[width=84mm]{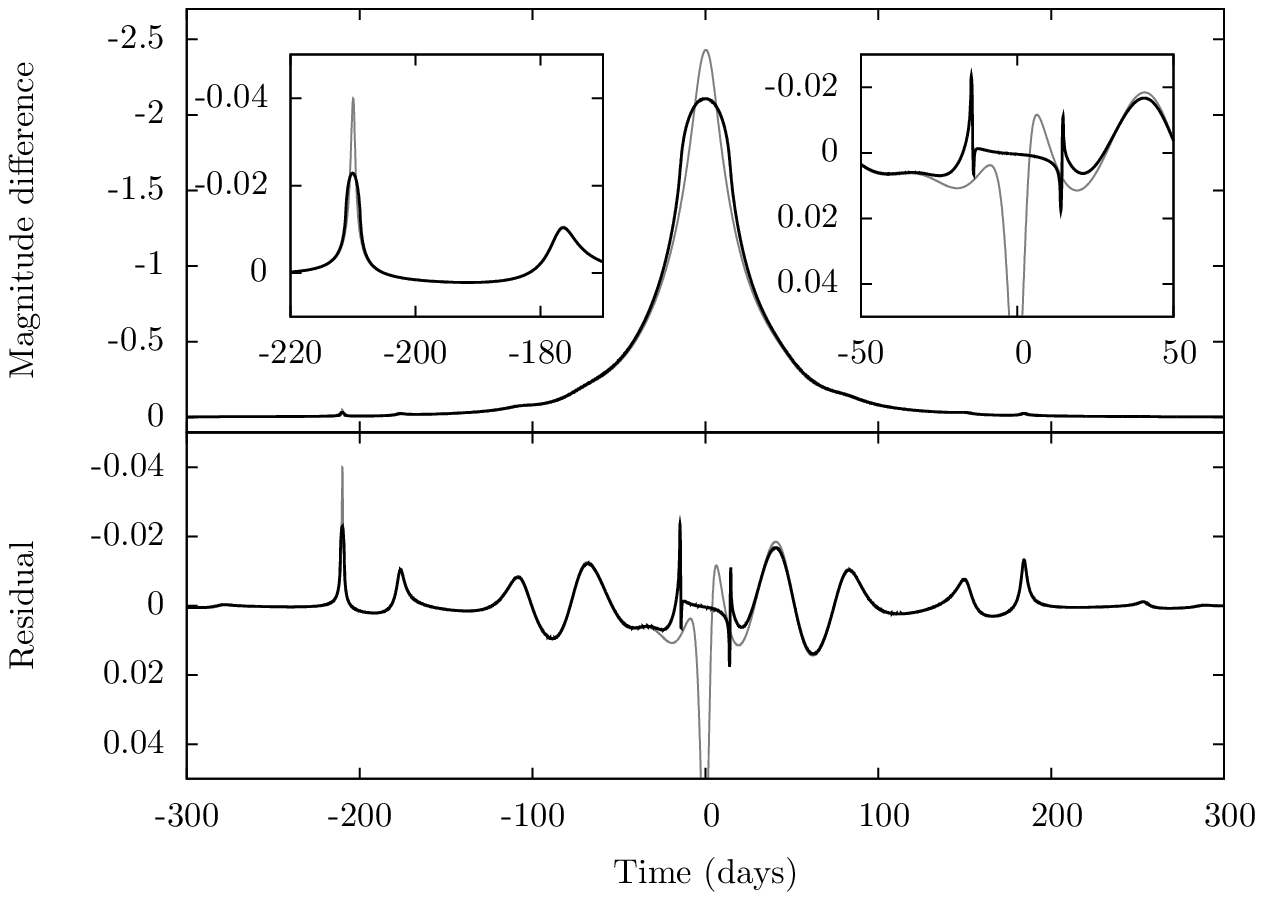}
\caption{The lightcurve of an RRL lensing a finite source of radius $100\rsun$
  (black) compared to the lightcurve of the same RRL lensing a point
  source (grey). The inset figures show in more detail the
  residuals when the source is close to the secondary caustic (on the
  left) and the central caustic (on the right). The lens has a mass $M=0.8\msun$, semimajor axis $a=0.4$~au, mass ratio $q=0.3$, fractional lens
  distance $x=0.95$, source distance $\ds=8$~kpc, source velocity
  $\vt=50$~\kms, impact parameter $\uzero=0.1$ and phase angle
  $\phizero=\upi/4$. The ratio of source to Einstein angular radii 
  $\rho_{\mathrm{s}}=\theta_{\mathrm{s}}/\thetae=0.28$ is very
  large. The effects of finite sources are only significant when the
  source is near the central or secondary caustics.}
\label{finiteSourceExampleLC}
\end{figure}

Figure~\ref{finiteSourceExampleLC} shows the lightcurve of an RRL
lensing a giant source of radius $100\rsun$, in comparison to the same
RRL lensing a point source. The effect of the finite source on the
lightcurve is clear, as it causes a wider, lower peak
magnification. Whilst the lens centre of mass transits the source,
there is effectively no deviation from the finite-source point-lens
lightcurve, except for spikes in the residual at $t \approx \pm 20$~d,
which are characteristic of a large source crossing a small central
caustic~\citep{Dong:2009bjp,Han:2009ccs}. In the wings of the lightcurve
there is very little difference between the finite- and point-source
lightcurves, and most of the features in the residuals have the same
amplitude. Only when the source is very close to the secondary caustic
is there any deviation from the point-source lightcurve in the
wings. The left inset of Figure~\ref{finiteSourceExampleLC} shows that
the peak in the finite-source lightcurve at $t \approx -210$~d is
slightly broader and about half the amplitude of the point-source
lightcurve. Interestingly, this peak, although broadened by the finite-source, is still much narrower than the source crossing time, which
determines the width of the central peak. Its width is instead
determined by the time taken for the secondary caustic to cross a
source diameter.

The example we have shown is very extreme, with a very large source,
very close to the lens, and even then the finite-source effects only
render binary features undetectable over a relatively small fraction
of the lightcurve. A typical giant source star will be up to a factor
of ten smaller, so the part of the lightcurve severely affected by
finite-source effects will be correspondingly smaller. As the
source has to be transited by the lens centre for finite-source
effects to become apparent at the lightcurve peak, the probability of
this occurring is also reduced by the same factor. This means that
finite-source effects will not affect the detectability of repeating
features very much. If finite-source effects are detected in an event,
the measurement of the source radius, combined with a measurement of
the lens period can be used together to measure the lens mass to a two
fold degeneracy~\citep{Dominik:1998mrb}. 

\subsection{Inclination and eccentricity}
\label{inclinationAndEccentricity}

\begin{figure}
\includegraphics[width=84mm]{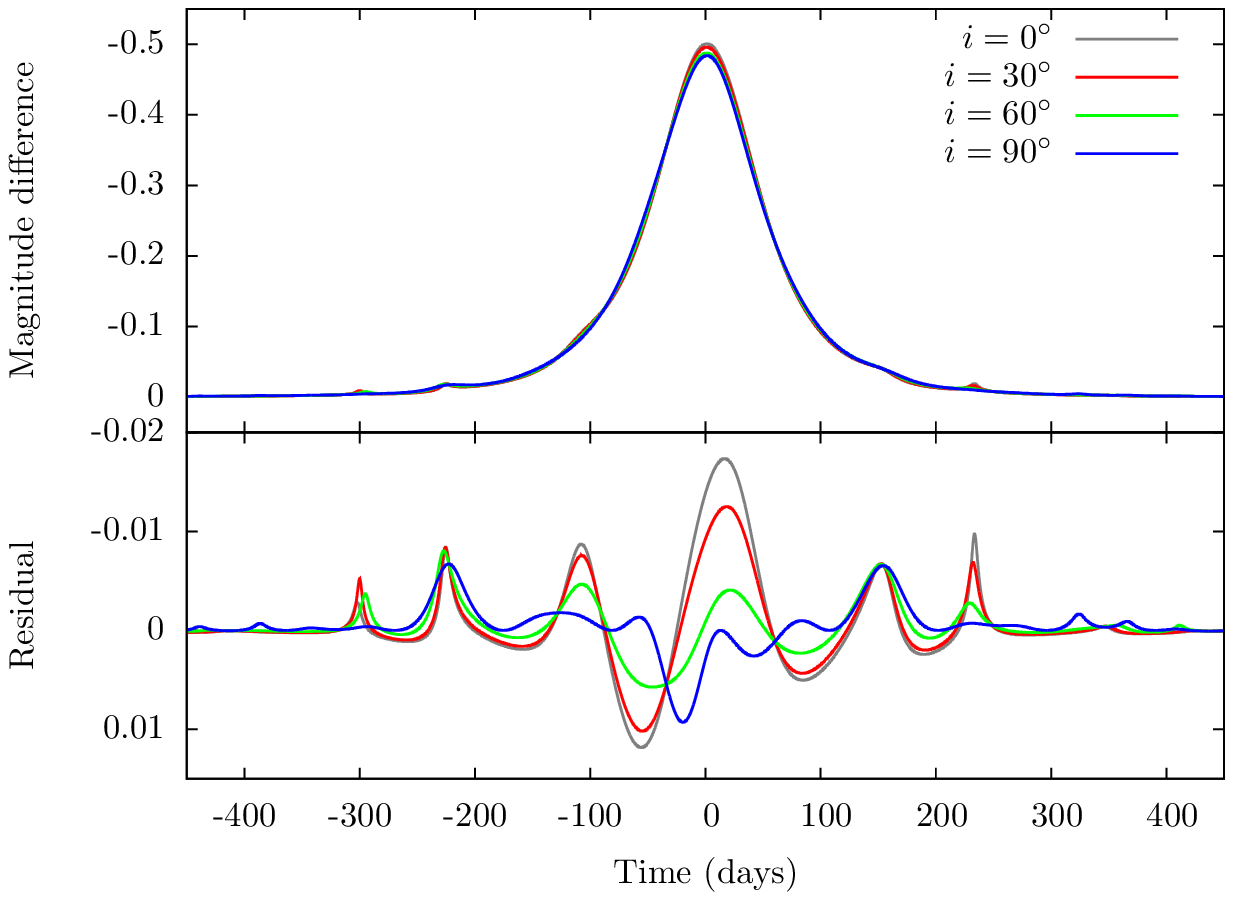}
\caption{The lightcurves of RRLs with different orbital inclinations
  relative to the line of sight. For each lightcurve, the lens has
  mass $M=0.58\msun$, semimajor axis $a=0.54$~au, mass ratio $q=0.52$,
  fractional lens distance $x=0.86$, source distance $\ds=9.5$~kpc,
  source velocity $\vt=61$~\kms, impact parameter $\uzero=0.77$, phase
  angle $\phizero=4.3$ measured in the plane of the orbit. The orbit
  was circular, and inclined about the $x$-axis as defined in
  Figure~\ref{parametrization}.}
\label{inclinationExampleLC}
\end{figure}

Inclination and eccentricity of the lens orbit will act to make the
magnification pattern motion much more complicated, as changes in the
projected lens separation cause the caustics to move and change
shape~\citep[see e.g. Figures~21 and 22 of][]{Penny:2011omm}. The effects are too complicated to investigate in detail, but
it is worth considering them in brief. For a lens with a given
semimajor axis, inclining the orbit should reduce the detectability of
features over part of the orbit, as $s$
decreases. Figure~\ref{inclinationExampleLC} shows the effect of
inclination on the lightcurve of an RRL. It shows that inclination
tends to decrease the amplitude of features, but does not completely
wipe them out, even when the inclination $i=90\degr$. In this extreme
case, rather than rotating, the secondary caustics move along diagonal
lines as the projected separation of the lenses changes, but their
angle does not, except for flips by $\upi$ every half
period. Inclination significantly changes the morphology of the
lightcurve, and can also change the timing of peaks (see e.g. those at
$t \approx -300$~d), which implies that it may be possible to measure
the inclination of the lens orbit from the lightcurve.

In contrast to inclination, eccentricity may increase the
detectability of features. For a given semimajor axis, eccentricity
can both increase and decrease the projected separation. However,
Kepler's second law implies that the lens will spend longer at larger
projected separations (assuming no inclination). As with inclination,
eccentricity will also change the lightcurve morphology and timing
of features, so it may also be possible to measure the eccentricity of
the lens from the lightcurve. Simultaneously including the effects of
inclination and eccentricity in the modelling of an RRL event will
likely be difficult, as together they require an additional four
parameters over the standard RRL parametrization. However, as the
angle of magnification pattern features does not depend strongly on
the projected separation, it will be possible to include inclination
and eccentricity in the timing analysis proposed in
Section~\ref{physicalParameters}, which may significantly ease the
analysis by narrowing down the search space to the range of parameters
compatible with timing measurements.

\subsection{Parallax}
\label{parallax}

Parallax effects due to the motion of the Earth about the Sun will
cause the source to appear to take a curved path through the plane of
the sky, and will affect the lightcurve of an RRL event. If the magnitude
of the parallax effect is small then it will cause only small
perturbations to the shape of the lightcurve and the timing of
features. Larger effects may cause significant changes to the RRL
lightcurve, significantly changing the timing of features, and
possibly making them appear less periodic, or adding a stronger annual
periodicity to the lightcurve. However, while parallax may
significantly complicate the interpretation of an RRL event, it does
not affect the magnification map, and the detectability of RRL
features should remain the same. Moreover, the detection of parallax
in an RRL event will allow the lens mass to be measured, at least to a
two-fold degeneracy (see Section~\ref{physicalParameters}). Due to the
photometric accuracy required to detect RRLs, and the long timescales of the
events, the probability of detecting parallax along with RRL features
is significant~\citep{Buchalter:1997pme}.

\end{document}